\documentclass[10pt]{article}

\usepackage{amsmath,amsfonts}
\usepackage{amssymb}
\usepackage{graphicx}

\newcommand{\pa}{\partial}

\newcommand{\be}{\begin{equation}}
\newcommand{\ee}{\end{equation}}

\begin{document}

\title{Coupled complex Ginzburg-Landau systems with saturable nonlinearity and asymmetric cross-phase modulation}

\author{Robert A. Van Gorder$^*$, Andrew L. Krause, Ferran Brosa Planella, and Abigail M. Burton \\  \small  Mathematical Institute, University of Oxford\\ \small Andrew Wiles Building, Radcliffe Observatory Quarter, Woodstock Road \\ \small Oxford, OX2 6GG, United Kingdom\\
\small Email: Robert.VanGorder@maths.ox.ac.uk}        
\date{\today}

\maketitle

\begin{center}
Abstract
\end{center}
We formulate and study dynamics from a complex Ginzburg-Landau system with saturable nonlinearity, including asymmetric cross-phase modulation (XPM) parameters. Such equations can model phenomena described by complex Ginzburg-Landau systems under the added assumption of saturable media. When the saturation parameter is set to zero, we recover a general complex cubic Ginzburg-Landau system with XPM. We first derive conditions for the existence of bounded dynamics, approximating the absorbing set for solutions. We use this to then determine conditions for amplitude death of a single wavefunction. We also construct exact plane wave solutions, and determine conditions for their modulational instability. In a degenerate limit where dispersion and nonlinearity balance, we reduce our system to a saturable nonlinear Schr\"odinger system with XPM parameters, and we demonstrate the existence and behavior of spatially heterogeneous stationary solutions in this limit. Using numerical simulations we verify the aforementioned analytical results, while also demonstrating other interesting emergent features of the dynamics, such as spatiotemporal chaos in the presence of modulational instability. In other regimes, coherent patterns including uniform states or banded structures arise, corresponding to certain stable stationary states. For sufficiently large yet equal XPM parameters, we observe a segregation of wavefunctions into different regions of the spatial domain, while when XPM parameters are large and take different values, one wavefunction may decay to zero in finite time over the spatial domain (in agreement with the amplitude death predicted analytically). We also find a collection of transient features, including transient defects and what appear to be rogue waves, while in two spatial dimensions we observe highly localized pattern formation. While saturation will often regularize the dynamics, such transient dynamics can still be observed - and in some cases even prolonged - as the saturability of the media is increased, as the saturation may act to slow the timescale.

\section{Introduction}

The complex Ginzburg-Landau (GL) equation models a variety of phenomena, including nonlinear waves, second-order phase transitions, superconductivity, superfluidity, Bose-Einstein condensation, liquid crystals, and strings in field theory \cite{aranson2002world}. The most common example is the cubic GL equation, for which a variety of solutions have been found \cite{nozaki1984exact}. Spatiotemporal chaos has been observed in the complex cubic GL equation \cite{shraiman1992spatiotemporal,chate1994spatiotemporal}. Transition to spatiotemporal turbulence from traveling waves due to instabilities was discussed in \cite{weber1992stability}. Spatial patterning is also possible, with solutions such as spiral waves being found \cite{biktasheva1998localized}. Bounds on the attractor for complex cubic GL equations have been considered in \cite{mielke1997complex,mielke1998bounds}. Quintic or cubic-quintic GL equations have been studied \cite{afanasjev1996three}, and have application to areas of nonlinear optics including lasers \cite{moores1993ginzburg}, optical waveguides equipped with a Bragg grating \cite{atai2001families}, quasi-CW Raman fiber lasers \cite{sugavanam2015ginzburg}, and Bose-Einstein condensates \cite{pinsker2015approximate,luckins2018bose}. Solutions to non-autonomous forms of the complex GL equation also have been studied \cite{wong2015dromion}. 

Coupled or vector complex cubic GL equations have also attracted interest. For example, the effect of slow real modes in reaction-diffusion systems close to a supercritical Hopf bifurcation was studied in \cite{ipsen2000finite} via a system of two coupled GL equations, where it was pointed out that a single GL equation cannot reproduce the dynamics from the reaction-diffusion system, suggesting the necessity of considering such coupled systems in some contexts; see also \cite{neufeld1996noise}. Such systems arise in various applications in nonlinear optics \cite{nistazakis2002multichannel}.  In the case where the squared-modulus of the waves feature into each equation (in contrast, say, to the case of linear coupling \cite{malomed2007solitary}), the cross-phase modulation parameters may vary \cite{liu2013elastic}. This is modeled by assuming different coefficients for the self- and cross-interaction terms, as was done in \cite{sakaguchi1995phase}. Phase instabilities in vector GL equations have been studied \cite{san1995phase}. A coupled GL system formulation of the weak electrolyte model for electroconvection in planarly aligned nematic liquid crystals has been considered \cite{treiber1998coupled}. A coupled complex GL system for the amplitudes of waves along a gas-less combustion front was derived in \cite{matkowsky1992coupled}. 

A variety of solutions to systems of coupled cubic GL equations have been studied in the literature \cite{matkowsky1993stability, sakaguchi1995phase}. Instability and transition from chaotic traveling waves to stable states such as soliton lattices was studied in \cite{sakaguchi1996localized,sakaguchi1996traveling}. Spatiotemporal chaos in coupled complex GL equations was studied in \cite{hernandez1999spatiotemporal}. Creation and annihilation processes of different kinds of vector defects in coupled GL equations are considered in \cite{hoyuelos2003dynamics}, as are transitions between different regimes of spatiotemporal dynamics. Phase singularities in such systems are studied in \cite{hoyuelos2005characterization}.

Although far less studied than their cubic counterparts, coupled cubic-quintic GL equations have also been studied in the literature. Modulational instability of linearly coupled equations was considered in \cite{porsezian2009modulational}. The collision of pulses or of solitary waves has also attracted attention \cite{descalzi2006collisions,descalzi2007collisions,descalzi2009noise}. The spatiotemporal structure of pulsating solitons was approximated using the variational approximation \cite{mancas2009spatiotemporal}. Modulational instability of plane waves in two coupled cubic-quintic GL equations coupled with a XPM term (in the cubic nonlinearity alone) was considered in \cite{alcaraz2010modulational} and also \cite{zakeri2015modulational}, while the dynamics of stationary pulse solutions for such equations were studied in \cite{alcaraz2011stationary}. Synchronization of solutions in such systems was studied in \cite{ciszak2015anticipated}.

Polynomial nonlinearities are not the only possibility for self- or cross-interaction terms, however cubic or cubic-quintic nonlinearity is what is standard in the literature. In the case of nonlinear Schr\"odinger equations, a variety of other self-interaction terms are often studied. Saturable nonlinearities consist of a rational or other bounded function of the complex modulus of the macroscopic wave function \cite{jakubowski1997information,litchinitser1999asymmetric}. These find use in nonlinear optics \cite{marburger1968dynamical, weilnau2002spatial}, with application to nematic liquid crystals \cite{reinbert2006spatial,skuse2008two}, optically induced nonlinear photonic lattices \cite{fleischer2003observation}, and have been used to study left-handed properties of metamaterials with saturable nonlinearity \cite{maluckov2008left}, to name a few applications. Despite this interest in saturable nonlinearities, there do not appear to be vector or even scalar extensions of complex GL equations which employ such nonlinearity, although such systems are natural generalizations of saturable NLS systems to the dissipative regime. This extension shall therefore be the focus of the present paper.

The remainder of this paper is organized as follows. In Section \ref{system}, we formulate a coupled complex GL system with saturable nonlinearity and XPM parameters in the nonlinear terms. In Section \ref{analysis}, we study the system analytically, approximating the absorbing set for the solution trajectories. One interesting phenomena observed is extinction of one of the wavefunctions. This depends strongly on the XPM parameters - and we arrive at local conditions for this to occur for either $u$ or $v$. In Section \ref{modinst}, we consider the modulational instability of the saturable coupled complex GL system, demonstrating instability of planar waves under small wavenumber perturbations. In Section \ref{sec2e}, we consider a singular limit which corresponds to an integrable reduction of the (in general, non-integrable) system, obtaining a variety of stationary solutions. In Section \ref{numeric}, we verify the aforementioned analytical results numerically, while also demonstrating other features of the system dynamics in non-integrable regimes. As suggested analytically, there exists an absorbing set for sufficiently small $\epsilon$, otherwise the solutions blow-up for large time. Spatiotemporal chaos is common, although there are some parameter regimes where coherent patterns - such as uniform states or banded structures - appear to persist. A partitioning of the support of wave functions, with the wave functions tending to separate into distinct regions over time, is observed in many cases. In the extreme case, one wavefunction will decay to zero with the other persisting, as the local analytical results suggested. In addition to these global behaviors, we observe a variety of highly local structures, and while spatiotemporal chaos appears quite common, we note for some parameter regimes the appearance of transient defects and what appear to be rogue waves at intermediate time intervals. The primary role of the saturation appears to be regularization of the cubic GL dynamics and a modification of the timescale of the dynamics, depending on the parameter regime, and the kinds of dynamics we observe can all be observed in cubic GL systems with XPM parameters (save for blow-up or high-intensity solutions which are particular to the saturable model). We summarize the interesting findings and discuss our results in Section \ref{dis}.

\section{Coupled complex GL system with saturable nonlinearity}\label{system}
To the best of our knowledge, saturable nonlinearity has not been considered in coupled complex GL dynamics, and this shall be the focus of our paper. In this section, we shall begin with the saturable coupled NLS system, and then discuss its generalization to the complex GL regime. Although complex GL equations and systems often are derived as modulational or amplitude equations for more complicated models describing specific physical phenomena (for which real GL equations and systems are a distinguished limit), complex GL equatios and systems can alternatively be obtained as dissipative perturbations of corresponding NLS equations or systems (for which the distinguished limit is the NLS equation or system) \cite{aranson2002world}. It is the latter regime which shall motivate our use of saturable nonlinearities in complex GL models. 

Saturable NLS systems commonly take the form
\be \label{nls1}
i\frac{\pa u}{\pa t} = \tilde{a}\nabla^2 u  + \frac{\tilde{b}\left(|u|^2 + \xi_1 |v|^2\right)u}{\left\lbrace 1+ \mu \left( |u|^2 + |v|^2\right)^n\right\rbrace^{1/n} }\,,
\ee
\be \label{nls2} 
i\frac{\pa v}{\pa t} = \tilde{a}\nabla^2 v + \frac{\tilde{b}\left(\xi_2 |u|^2 + |v|^2\right)v}{\left\lbrace 1+ \mu \left( |u|^2 + |v|^2\right)^n\right\rbrace^{1/n} }\,,
\ee
where $\tilde{a}$ and $\tilde{b}$ give the relative strengths of the dispersion and nonlinearity, respectively, while $\xi_1$ and $\xi_2$ are the XPM parameters (self-interaction parameters are scaled to unity). We take the Laplacian operators $\nabla^2$ to be defined over spatial domains of the form $\mathbb{R}^{k}$, $k = 1,2,\dots$. The saturable $n=1$ case was considered in \cite{jakubowski1997information,litchinitser1999asymmetric,fleischer2003observation,maluckov2008left}. The $n=1$ case gives a simple rational function, and has physical relevance to modeling optical beams in photorefractive media \cite{christodoulides1996incoherently, kutuzov1998cross, musslimani1999suppression, malmberg2000vector, belic2003self, salgueiro2004single, chow2006exact}. The $n=2$ case was considered in \cite{reinbert2006spatial,skuse2008two}, with applications to nematic liquid crystals discussed. In addition to nonlinear optics applications, the algebraic nonlinearity (corresponding to $n=2$, but of a different form) has been suggested as a model of BEC dynamics under tight transverse confinement \cite{salasnich2002effective, chow2006exact}. In addition to saturable NLS systems on continuum spatial domains, there also exist models of saturable NLS dynamics on spatial lattices; see \cite{conte2008doubly}.

While a variety of numerical and some analytical solutions have been reported in the above references, a fairly systematic study of exact solutions to systems of the form \eqref{nls1}-\eqref{nls2} for $n=1$ were reported in \cite{chow2006exact}. In particular, XPM terms distinct from self-interaction terms were considered in \cite{chow2006exact} for the case of $n=1$ (and also for a different form of the equations when $n=2$). A number of exact soliton solutions were obtained when the equations were integrable, and these solutions were obtained under the assumption that the XPM parameters were equal yet distinct from unity, i.e. $\xi_1 = \xi_2 \neq 1$. However, in order for many of the solutions to exist with $\xi_1 = \xi_2 \neq 1$, distinct saturable parameters were needed in each equation. There are also existence results for certain NLS systems of the form \eqref{nls1}-\eqref{nls2}, particularly for $n=1$ \cite{de2013weakly, lin2014ground}. See, also, \cite{lin2017virial} for energy estimates and a virial theorem. 

Hamiltonian and Lagrangian formulations for scalar saturable NLS equations \cite{karlsson1992optical, gatz1997propagation} and coupled saturable NLS systems \cite{skuse2008two, skuse2009interaction, lin2014ground, cao2018normalized} have been considered in the literature. Following the perturbation structure outlined in works such as \cite{pereira1977nonlinear, weiland1978perturbation, cruz2004complex}, one may perturb the Hamiltonian form of the system \eqref{nls1}-\eqref{nls2} so that there are dissipative dispersion and nonlinearity contributions. In the cubic NLS case, these dissipative terms may be viewed as modeling growth and damping of a wave under NLS in a more realistic media \cite{pereira1977nonlinear,weiland1978perturbation}, and this interpretation carries over to the saturable NLS case, as well. Such perturbation approaches have resulted in complex GL equations which were originally obtained to stdy nonlinear Landau damping \cite{goldman1977dissipative, weiland1978perturbation}. As pointed out by \cite{elphick1989localized, elphick1990comment, van1992fronts}, the NLS structures possess dilational and Galilean symmetries which are then destroyed under these generic dissipative perturbations.

Obtaining the Euler-Lagrange equations of motion for the perturbed saturable NLS Hamiltonian, and appropriately rescaling all quantities in a manner which is consistent with the standard canonical form for the cubic GL system, one obtains a saturable coupled complex GL system taking the form 
\be \label{sgl1}
\frac{\pa u}{\pa t} = \epsilon u + \left( 1 + ia\right)\nabla^2 u - \left( 1 - ib\right) \frac{\left(|u|^2 + \alpha_1 |v|^2\right)u}{\left\lbrace 1+ \mu \left( |u|^2 + |v|^2\right)^n\right\rbrace^{1/n} }\,,
\ee
\be \label{sgl2} 
\frac{\pa v}{\pa t} = \epsilon v + \left( 1 + ia\right)\nabla^2 v - \left( 1 - ib\right) \frac{\left(\alpha_2 |u|^2 + |v|^2\right)v}{\left\lbrace 1+ \mu \left( |u|^2 + |v|^2\right)^n\right\rbrace^{1/n} }\,,
\ee
where $\epsilon, a, b \in \mathbb{R}$, $\mu >0$, $n=1,2,\dots$, and $\alpha_1,\alpha_2$ are XPM parameters. Both $\mu$ and $n$ determine respectively the strength and type of saturable term. The parameter $\epsilon$ encodes the linear Landau damping (or, gain, depending on sign), while the real coefficient of the nonlinear term (which is equal to unity after scaling) can be viewed as a nonlinear damping. 

The interpretation of the terms present in \eqref{sgl1}-\eqref{sgl2} carries over from the complex GL case \cite{van1992fronts}, with the modification being the inclusion of a saturable nonlinearity. Indeed, observe that when $\mu=0$, we recover the well-studied cubic complex GL system
\be \label{cgl1}
\frac{\pa u}{\pa t} = \epsilon u + \left( 1 + ia\right)\nabla^2 u - \left( 1 - ib\right) \left(|u|^2 + \alpha_1 |v|^2\right)u\,,
\ee
\be \label{cgl2}
\frac{\pa v}{\pa t} = \epsilon v + \left( 1 + ia\right)\nabla^2 v - \left( 1 - ib\right) \left(\alpha_2 |u|^2 + |v|^2\right)v\,,
\ee
and hence we expect that dynamics between the saturable model \eqref{sgl1}-\eqref{sgl2} and the cubic model \eqref{cgl1}-\eqref{cgl2} will be similar in the regime where saturable effects are more negligible, such as the small-amplitude regime. Meanwhile, when the dynamics are of large amplitude and strongly nonlinear, we anticipate certain differences between the two models, which shall be discussed throughout the present paper.

Regarding XPM terms in cubic GL systems, note that the literature consists of papers which take $\alpha_1 = \alpha_2$, and often both are set to unity. The case of different XPM parameters has apparently not been considered under cubic GL dynamics (coupled complex cubic \cite{nistazakis2002multichannel} or cubic-quintic \cite{alcaraz2010modulational,alcaraz2011stationary,zakeri2015modulational} GL equations coupled with the two XPM parameters always equal have been considered  but with very specific solutions sought and generic dynamics not explored, while coupled complex GL equations with two distinct XPM parameters do not appear to have been written down let alone explored). As we shall see, interesting dynamics will arise from \eqref{sgl1}-\eqref{sgl2} when $\alpha_1 \neq \alpha_2$, even when $\mu =0$ (corresponding to the cubic GL system \eqref{cgl1}-\eqref{cgl2}). This suggests that even in the $\mu=0$ limit, this system may give new dynamics for some combinations of XPM parameters.

\section{Asymptotic dynamics of saturable complex GL systems}\label{analysis}
In the present section we shall study the asymptotic dynamics of \eqref{sgl1}-\eqref{sgl2} under certain assumptions which shall permit bounded solutions. We first construct an approximation to the absorbing set for solution trajectories in $(u,v)\in \mathbb{C}^2$, and later numerical simulations validate the results we obtain despite the fact that these results are asymptotic in nature. Simulations suggest that the dynamics of \eqref{sgl1}-\eqref{sgl2} are often modulationally unstable, and we are able to demonstrate that there always exist unstable wavenumbers for generic parameter values which induce modulational instability. One interesting feature observed in numerical simulations is the dissipation or extinction of one wavefunction under certain restrictions on the XPM parameters, and we are able to determine these conditions analytically. Finally, all of our results for the system \eqref{sgl1}-\eqref{sgl2} can be deduced for the scalar saturable GL equation, and we list those results for completeness.

\subsection{Approximating the absorbing set for the system \eqref{sgl1}-\eqref{sgl2}}\label{sec3.1}
In order to construct an approximation to the absorbing set for solutions in $\mathbb{C}^2$, we shall make the simplifying assumption that there is one dominant spatial mode for sake of analytical tractability. Assuming that arbitrarily small modes are excited on the plane, we shall then take the wavenumbers corresponding to dominant modes to be arbitrarily small. While this is not generically true, we note that asymptotic results obtained under this assumption show agreement with results from numerical simulations (which shall be considered later in Sec. \ref{numeric}). One benefit to our approach is that we shall be able to approximate the dynamics on the absorbing set via a system of ordinary differential equations. The resulting planar dynamics can be classified, and will permit us to determine specific parameter conditions for bounded yet non-trivial dynamics.

We shall follow the mathematical approach of \cite{van2018generic} to obtain an absorbing set for the dynamics. Our approach will differ in that the saturable nonlinearities permit explicit computations to be carried out, and we shall be able to give more concrete results in terms of the nonlinearities and their derivatives. In contrast, the general approach of \cite{van2018generic} can be used for a wide variety of nonlinear terms, and hence the results there often invoke inverse functions rather than derivatives. This will account for various differences in the derivations, rather than a straightforward application of the results in \cite{van2018generic}.

Consider a solution of the form
\be \label{tranf}
u = e^{i\mathbf{k}_u\cdot \mathbf{x}}U(t) \quad \text{and} \quad v = e^{i\mathbf{k}_v\cdot \mathbf{x}}V(t)\,.
\ee
Such a solution will give the dynamics corresponding to a single wavenumber. The transformation \eqref{tranf} puts \eqref{sgl1}-\eqref{sgl2} into the form
\be \label{T1}
\frac{dU}{dt} = \epsilon U - \left( 1 + ia\right)|\mathbf{k}_u|^2 U - \left( 1 - ib\right) \frac{\left(|U|^2 + \alpha_1 |V|^2\right)U}{\left\lbrace 1+ \mu \left( |U|^2 + |V|^2\right)^n\right\rbrace^{1/n} }\,,
\ee
\be \label{T2} 
\frac{dV}{dt} = \epsilon V - \left( 1 + ia\right)|\mathbf{k}_v|^2 V - \left( 1 - ib\right) \frac{\left(\alpha_2 |U|^2 + |V|^2\right)V}{\left\lbrace 1+ \mu \left( |U|^2 + |V|^2\right)^n\right\rbrace^{1/n} }\,.
\ee
Consider $U(t) = \rho_u(t)\exp(i\theta_u(t))$ and $V(t)=\rho_v(t)\exp(i\theta_v(t))$, which puts \eqref{T1}-\eqref{T2} into the form
\be\label{rhou} 
\frac{d\rho_u}{dt} = \left(\epsilon - |\mathbf{k}_u|^2\right)\rho_u - \frac{\left( \rho_u^2 + \alpha_1 \rho_v^2 \right)\rho_u}{\left\lbrace 1+ \mu \left( \rho_u^2 + \rho_v^2 \right)^n\right\rbrace^{1/n} }\,,
\ee
\be 
\rho_u\frac{d\theta_u}{dt} = - a|\mathbf{k}_u|^2 \rho_u + \frac{b\left( \rho_u^2 + \alpha_1 \rho_v^2 \right)\rho_u}{\left\lbrace 1+ \mu \left( \rho_u^2 + \rho_v^2 \right)^n\right\rbrace^{1/n} }\,,
\ee
\be \label{rhov}
\frac{d\rho_v}{dt} = \left(\epsilon - |\mathbf{k}_v|^2\right)\rho_v - \frac{\left( \alpha_2 \rho_u^2 + \rho_v^2 \right)\rho_v}{\left\lbrace 1+ \mu \left( \rho_u^2 + \rho_v^2 \right)^n\right\rbrace^{1/n} }\,,
\ee
\be 
\rho_v\frac{d\theta_v}{dt} = - a|\mathbf{k}_v|^2 \rho_v + \frac{b\left( \alpha_2\rho_u^2 + \rho_v^2 \right)\rho_v}{\left\lbrace 1+ \mu \left( \rho_u^2 + \rho_v^2 \right)^n\right\rbrace^{1/n} }\,.
\ee
Observe that we may write
\be 
\rho_u\frac{d\theta_u}{dt} = - a|\mathbf{k}_u|^2 \rho_u + b\left\lbrace \left( \epsilon - |\mathbf{k}_u|^2 \right)\rho_u - \frac{d\rho_u}{dt} \right\rbrace\,,
\ee
and solving for $\theta_u$ (and neglecting the integration constant, since it only provides a phase shift) we obtain
\be 
\theta_u(t) = \left\lbrace \epsilon b - (a+b)|\mathbf{k}_u|^2 \right\rbrace t - b \log |\rho_u(t)|\,. 
\ee
Similarly, 
\be 
\theta_v(t) = \left\lbrace \epsilon b - (a+b)|\mathbf{k}_v|^2 \right\rbrace t - b \log |\rho_v(t)|\,. 
\ee
This then leaves us with two coupled ODEs for $\rho_u$ and $\rho_v$, \eqref{rhou} and \eqref{rhov}.

Define the intensity 
\be \label{intensity}
I  = |u|^2+|v|^2\,.
\ee
We shall observe that the intensity \eqref{intensity} behaves as a Lyapunov function. Note that $I(0,0)=0$, while $I\left( |u|^2,|v|^2\right)>0$ for all $\left( |u|^2,|v|^2\right)\neq (0,0)$, and it is radially unbounded.
Note that 
\be \label{dIdt}
\frac{1}{2}\frac{dI}{dt} = \left( \epsilon - |\mathbf{k}_u|^2 \right)\rho_u^2 + \left( \epsilon - |\mathbf{k}_v|^2 \right)\rho_v^2 - \frac{\rho_u^4 + \rho_v^4 + (\alpha_1 + \alpha_2)\rho_u^2\rho_v^2}{\left\lbrace 1+ \mu \left( \rho_u^2 + \rho_v^2 \right)^n\right\rbrace^{1/n}}\,.
\ee
As $I$ is equivalent to the Euclidean norm on $\mathbb{R}^4$, we have four-volume contraction when $\frac{dI}{dt}<0$ and expansion when $\frac{dI}{dt} >0$. 

To better understand the structure of \eqref{dIdt}, write $\rho_u = r\cos(\phi)$ and $\rho_v = r\sin(\phi)$ so that the $\rho_u$-$\rho_v$ plane is put into polar coordinates. This puts \eqref{dIdt} into the form
\be 
\frac{1}{2}\frac{dI}{dt} = F\left( r^2, \sin^2(\phi) \right)\,,
\ee
where
\be \label{F}
F(s,\chi) = \left( \epsilon - |\mathbf{k}_u|^2(1-\chi) - |\mathbf{k}_v|^2 \chi \right) s - \frac{s^2\left(1 + (\alpha_1 + \alpha_2 -2)\chi(1-\chi) \right)}{\left\lbrace 1 + \mu s^n \right\rbrace^{1/n}}\,.
\ee
Here $F$ has domain $s\geq 0$ and $\chi \in [0,1]$. 

We shall only be interested in solutions for which $I(t)$ remains bounded - that is, solutions with bounded modulus for all time (given appropriate initial data). Therefore, we require that $F<0$ as $s\rightarrow \infty$. From \eqref{F}, we have
\be 
F(s,\chi) = \left\lbrace \epsilon - |\mathbf{k}_u|^2(1-\chi) - |\mathbf{k}_v|^2 \chi  - \frac{1 + (\alpha_1 + \alpha_2 -2)\chi(1-\chi) }{\mu^{1/n}} \right\rbrace s + O\left( s^{1-n} \right)\,.
\ee
Since $n\geq 1$, we have that $F \rightarrow -\infty$ as $s\rightarrow \infty$ for 
\be 
\epsilon < |\mathbf{k}_u|^2(1-\chi) + |\mathbf{k}_v|^2 \chi  + \frac{1 + (\alpha_1 + \alpha_2 -2)\chi(1-\chi) }{\mu^{1/n}}\,.
\ee
Now, we desire solutions which have bounded modulus for \textit{arbitrary} wavenumbers, so we set $|\mathbf{k}_u|=|\mathbf{k}_v |=0$ as these will include the least stable wavenumbers. With this, we have
\be 
\epsilon < \frac{1 + (\alpha_1 + \alpha_2 -2)\chi(1-\chi) }{\mu^{1/n}}\,.
\ee
There are two cases to consider. First, if $\alpha_1 + \alpha_2 \geq 2$, then
\be 
\epsilon < \frac{1}{\mu^{1/n}}\,,
\ee
as we must maintain stability independent of the angle taken toward $s\rightarrow \infty$ in the plane. On the other hand, when $\alpha_1 + \alpha_2 < 2$, we have
\be 
\epsilon < \frac{1}{\mu^{1/n}}\left( 1 + \frac{\alpha_1+\alpha_2-2}{4}\right) = \frac{2+\alpha_1+\alpha_2}{4\mu^{1/n}}
\ee
as for this case the least stable trajectories as $s\rightarrow \infty$ correspond to $\chi = \frac{1}{2}$. Therefore, we have obtained the condition $\epsilon < \overline{\epsilon}(\alpha_1,\alpha_2,\mu,n)$ such that $F \rightarrow - \infty$ as $s\rightarrow \infty$, 
where we define
\be 
\overline{\epsilon}(\alpha_1,\alpha_2,\mu,n) = \begin{cases}
\frac{1}{\mu^{1/n}}, & \text{for} ~~ \alpha_1 + \alpha_2 \geq 2\,,\\
\frac{2+\alpha_1+\alpha_2}{4\mu^{1/n}}, & \text{for} ~~ \alpha_1 + \alpha_2 < 2\,.
\end{cases}
\ee
One may show that $\epsilon > \overline{\epsilon}(\alpha_1,\alpha_2,\mu,n)$ will give $F \rightarrow \infty$ as $s\rightarrow \infty$ for at least some $\chi \in [0,1]$.

Let us next note that $F(0,\chi) =0$. Meanwhile,
\be 
\frac{\pa F}{\pa s}(0,\chi) = \epsilon - |\mathbf{k}_u|^2(1-\chi) - |\mathbf{k}_v|^2 \chi \,.
\ee
Then, when $\epsilon >0$ there exist arbitrarily small wavenumbers for which $F>0$ in a neighborhood of the origin. This, in turn, implies instability near the origin (instability of initial solutions with arbitrarily small modulus). On the other hand, if $\epsilon <0$, then the origin is locally stable. As we are interested in bounded yet nontrivial dynamics, we consider the case where $\epsilon >0$. 

The above analysis suggests that in order to obtain bounded yet non-uniform dynamics, we should consider $0 < \epsilon < \overline{\epsilon}(\alpha_1,\alpha_2,\mu,n)$. Now that we have this condition, we shall attempt to approximate the attractor. Note that our analysis presupposes that a fixed wavenumber will be excited, while in general multiple wavenumbers may be excited simultaneously. We shall therefore consider the case where the wavenumbers are zero, which is the least stable case to consider. This will then provide an approximation to the absorbing set, which can be viewed as being a bound for the attractor \cite{robinson2001infinite}. 

One may verify that for $\mu >0$, $n\geq 1$, $0 < \epsilon < \overline{\epsilon}(\alpha_1,\alpha_2,\mu,n)$, the function $\frac{\pa F}{\pa s}$ can have either zero or one turning points; either it decreases monotonically toward a fixed limit given by $\lim_{s\rightarrow \infty}F(s,\chi)/s$ or it deceases, then increases monotonically toward a fixed limit given by $\lim_{s\rightarrow \infty}F(s,\chi)/s$. By our choice of bounds on $\epsilon$, the limit will always be negative. Therefore, $\frac{\pa F}{\pa s} >0$ at $s=0$, $\lim_{s\rightarrow \infty}\frac{\pa F}{\pa s} <0$ , and since there is at most one turning point, for each $\chi\in [0,1]$, there will be a unique $s=s^{**}(\chi)$ at which $\frac{\pa F}{\pa s} =0$. This, in turn, suggests that along any fixed $\chi \in [0,1]$, there will exist a unique $s=s^*(\chi)$ at which $F$ is maximal (and, positive). Hence, along any fixed $\chi \in [0,1]$, we have that $F$ increases from zero to $F(s^*(\chi),\chi)$, and then decreases without bound ($F \rightarrow -\infty$ as $s\rightarrow \infty$). From this and continuity of $F$, there is a unique curve $s=\zeta(\chi)>0$ such that $F(\zeta(\chi),\chi)=0$.

To find the curve $\zeta(\chi)$, we set $F(\zeta,\chi)=0$ and, using $\zeta >0$, we have
\be 
\left( \epsilon - |\mathbf{k}_u|^2(1-\chi) - |\mathbf{k}_v|^2 \chi \right)  - \frac{\zeta\left(1 + (\alpha_1 + \alpha_2 -2)\chi(1-\chi) \right)}{\left\lbrace 1 + \mu \zeta^n \right\rbrace^{1/n}} =0\,,
\ee
which gives the relation
\be 
\frac{\zeta^n}{1+\mu \zeta^n} = \left\lbrace \frac{\epsilon - |\mathbf{k}_u|^2(1-\chi) - |\mathbf{k}_v|^2 \chi }{1 + (\alpha_1 + \alpha_2 -2)\chi(1-\chi) } \right\rbrace^n \,.
\ee
Let us again assume arbitrarily small wavenumbers are excited. Then, we have
\be \label{curve1}
\zeta(\chi) = \frac{\epsilon }{1 + (\alpha_1 + \alpha_2 -2)\chi(1-\chi) } \left\lbrace 1 - \mu \left( \frac{\epsilon }{1 + (\alpha_1 + \alpha_2 -2)\chi(1-\chi) }\right)^n \right\rbrace^{1/n}\,.
\ee
Returning to coordinates $(r,\phi)$, we have that $r^2 = \zeta\left( \sin^2(\phi)\right)$. Returning to original coordinates, we have the curve
\be \label{curve}
|U|^2 + |V|^2 = \zeta\left( \sin^2(\phi) \right)\,,
\ee
where $\phi = \tan^{-1}(|V|/|U|)$ provided $|U|\neq 0$ and $\phi = \frac{\pi}{2}$ if $|U|=0$. From the above analysis, the curve is closed and bounded away from the origin and away from infinity, given $\mu >0$, $n\geq 1$, $0 < \epsilon < \overline{\epsilon}(\alpha_1,\alpha_2,\mu,n)$. Putting this all together, let use define the sets 
\be \label{set1}
\mathcal{B}_i := \left\lbrace (U,V)\in \mathbb{C}^2 : |U|^2 + |V|^2 < \zeta\left( \sin^2(\phi) \right) \right\rbrace \,,
\ee
\be \label{set2}
\mathcal{A} := \left\lbrace (U,V)\in \mathbb{C}^2 : |U|^2 + |V|^2 = \zeta\left( \sin^2(\phi) \right) \right\rbrace \,,
\ee
and 
\be \label{set3}
\mathcal{B}_o := \left\lbrace (U,V)\in \mathbb{C}^2 : |U|^2 + |V|^2 > \zeta\left( \sin^2(\phi) \right) \right\rbrace  \,.
\ee
We have that $\frac{dI}{dt}>0$ on $\mathcal{B}_i$ and $\frac{dI}{dt}<0$ on $\mathcal{B}_o$. As $I$ is simply a scaling of the Euclidean norm, we therefore have that trajectories grow in norm on $\mathcal{B}_i$ while they contract in norm on $\mathcal{B}_o$, suggesting that as $t\rightarrow \infty$ we have that trajectories $(U,V)$ converge to the attractor $\mathcal{A}$.

While the curve given in terms of $\zeta$ may seem a bit abstract, note that we may bound $\zeta$ in the $|U|$-$|V|$ plane. Define constants
\be 
\zeta_1 = \frac{\epsilon}{\left\lbrace 1 - \mu \epsilon^n \right\rbrace^{1/n}}\,,
\ee
\be 
\zeta_2 = \frac{4\epsilon}{2+\alpha_1+\alpha_2}\left\lbrace 1 - \mu \left( \frac{4\epsilon}{2+\alpha_1+\alpha_2} \right)^n\right\rbrace^{-1/n}\,.
\ee
Then, the surface $\mathcal{A}\in \mathbb{C}^2$ is enclosed between a 3-sphere of radius $\sqrt{\min\left\lbrace \zeta_1, \zeta_2\right\rbrace}$ and a 3-sphere of radius $\sqrt{\max\left\lbrace \zeta_1, \zeta_2\right\rbrace}$. Which of $\zeta_1$ or $\zeta_2$ is greatest will depend on $\alpha_1 + \alpha_2 \gtrless 2$. This suggests that, when arbitrary small wavenumbers are excited, the large-time dynamics will be found within the region $\min\left\lbrace \zeta_1, \zeta_2\right\rbrace < |U|^2 + |V|^2 < \max\left\lbrace \zeta_1, \zeta_2\right\rbrace$. This should be more numerically feasible to detect than the explicit form of the attractor $\mathcal{A}$.

Note that, in the case where there is a minimal non-zero wavenumber excited (bounded away from zero), the general idea is the same, with the form of $\zeta$ including $|\mathbf{k}_u|$ and/or $|\mathbf{k}_v|$, as needed. Defining the curve $\hat{\zeta}(\chi)$ by
\be \label{curve2}
\zeta(\chi) = \frac{\epsilon - |\mathbf{k}_u|^2(1-\chi) - |\mathbf{k}_v|^2 \chi }{1 + (\alpha_1 + \alpha_2 -2)\chi(1-\chi) } \left\lbrace 1 - \mu \left( \frac{\epsilon - |\mathbf{k}_u|^2(1-\chi) - |\mathbf{k}_v|^2 \chi }{1 + (\alpha_1 + \alpha_2 -2)\chi(1-\chi) }\right)^n \right\rbrace^{1/n}
\ee
in place of the curve $\zeta(\chi)$ given in \eqref{curve1}, the sets $\mathcal{B}_i$, $\mathcal{A}$, and $\mathcal{B}_o$ can be defined as in \eqref{set1}-\eqref{set3}. Such a construction holds provided $\mathbf{k}_u$ and $\mathbf{k}_v$ are chosen such that $\epsilon - |\mathbf{k}_u|^2(1-\chi) - |\mathbf{k}_v|^2 \chi >0$ for all $\chi \in [0,1]$. This restriction is equivalent to $\max\left\lbrace |\mathbf{k}_u| , |\mathbf{k}_v| \right\rbrace < \sqrt{\epsilon}$, which defines the interior of a hypercube in wavenumber space.

\subsection{Local stability conditions for single wavefunction extinction}\label{secloal}
In numerical simulations we occasionally observe that one of the wavefunctions $u$ or $v$ tends to zero as $t\rightarrow \infty$. We shall now derive conditions under which one of the states $(u,v)=(0,v)$ or $(u,v)=(u,0)$ are locally stable. These conditions, in turn, will give parameter restrictions for which one of the two extinction states is stable. This case was studied for more generic GL systems in \cite{van2018generic}. Since the saturable nonlinearity takes a very specific form, we shall be able to obtain more analytically explicit results for the system we study. 

The XPM parameters will play a strong role in these conditions, with the results suggesting that situations where $\alpha_1 = \alpha_2 = 1$, which are commonly taken in the literature when studying cubic GL systems, do not result in this wavefunction extinction, while asymmetric parameter conditions can allow for dissipation of one wavefunction as $t\rightarrow \infty$. The analysis is similar to what was done in Sec. \ref{sec3.1}, only now we focus on the regime where one of the two underlying ordinary differential equations has a stable zero solution. The conditions for the stability of such a zero solution will be equivalent to those granting amplitude death of one of the wave functions in the original system \eqref{sgl1}-\eqref{sgl2}. In this way, we obtain analytical conditions on the extinction of a single wavefunction. These results are later verified through numerical simulations, in Sec. \ref{numeric}.

For large time we were able to show that solutions tend to the absorbing set $\mathcal{A}$, so let us consider a solution on the absorbing set, again assuming that $\mu >0$, $n\geq 1$, $0 < \epsilon < \overline{\epsilon}(\alpha_1,\alpha_2,\mu,n)$. Then, parameterizing the moduli of the solutions to \eqref{rhou} and \eqref{rhov} as $\rho_u = r\cos(\phi)$, $\rho_v = r\sin(\phi)$, we obtain
\be 
\frac{dr}{dt} = \left\lbrace \epsilon - |\mathbf{k}_u|^2 \cos^2(\phi) - |\mathbf{k}_v|^2 \sin^2(\phi) - \frac{\left( \cos^4(\phi) + \sin^4(\phi) + (\alpha_1+\alpha_2)\sin^2(\phi)\cos^2(\phi)\right)r^2}{\left\lbrace 1 + \mu r^{2n}\right\rbrace ^{1/n}} \right\rbrace r\,,
\ee
\be 
\frac{d\phi}{dt} = \left\lbrace |\mathbf{k}_u|^2 - |\mathbf{k}_v|^2 + \frac{\left( (1-\alpha_2)\cos^2(\phi) - (1-\alpha_1)\sin^2(\phi) \right)r^2}{\left\lbrace 1 + \mu r^{2n}\right\rbrace ^{1/n}} \right\rbrace  \sin(\phi)\cos(\phi)\,.
\ee
Now, on the absorbing set, the $r$ equation is satisfied for $r = \sqrt{\zeta\left( \sin^2(\phi)\right)}$, while the $\phi$ equation becomes
\be 
\begin{aligned}
\frac{d\phi}{dt} & = \left\lbrace |\mathbf{k}_u|^2 - |\mathbf{k}_v|^2 + \frac{\left( (1-\alpha_2)\cos^2(\phi) - (1-\alpha_1)\sin^2(\phi) \right)\zeta\left( \sin^2(\phi)\right)}{\left\lbrace 1 + \mu \left(\zeta\left( \sin^2(\phi)\right)\right)^{n}\right\rbrace ^{1/n}} \right\rbrace  \sin(\phi)\cos(\phi)\\
& = \left\lbrace |\mathbf{k}_u|^2 - |\mathbf{k}_v|^2 + \frac{\left[ (1-\alpha_2)\cos^2(\phi) - (1-\alpha_1)\sin^2(\phi)\right]\left[ \epsilon - |\mathbf{k}_u|^2 \cos^2(\phi) - |\mathbf{k}_v|^2 \sin^2(\phi) \right]}{1+\left( \alpha_1 + \alpha_2 - 2\right)\sin^2(\phi)\cos^2(\phi)} \right\rbrace  \sin(\phi)\cos(\phi)\\
& \equiv \Phi(\phi)\,.
\end{aligned}
\ee

We have that 
\be 
\frac{d\Phi}{d\phi}(\phi =0) = |\mathbf{k}_u|^2 - |\mathbf{k}_v|^2 + (1-\alpha_2)\left( \epsilon - |\mathbf{k}_u|^2\right)
\ee
and
\be 
\frac{d\Phi}{d\phi}\left(\phi =\frac{\pi}{2}\right) = -|\mathbf{k}_u|^2 + |\mathbf{k}_v|^2 + (1-\alpha_1)\left( \epsilon - |\mathbf{k}_v|^2\right)\,,
\ee
hence $\phi =0$ is locally stable if 
\be 
|\mathbf{k}_u|^2 - |\mathbf{k}_v|^2 + (1-\alpha_2)\left( \epsilon - |\mathbf{k}_u|^2\right) <0,
\ee 
while $\phi = \frac{\pi}{2}$ is locally stable if 
\be 
-|\mathbf{k}_u|^2 + |\mathbf{k}_v|^2 + (1-\alpha_1)\left( \epsilon - |\mathbf{k}_v|^2\right) <0.
\ee 

We arrive at the stability criteria for extinction of one of the wavefunctions. Given $\mu >0$, $n\geq 1$, $0 < \epsilon < \overline{\epsilon}(\alpha_1,\alpha_2,\mu,n)$, the state $(u,v)=(u,0)$ (corresponding to $\phi =0$) is locally stable on the absorbing set $\mathcal{A}$ provided that 
\be \label{die1}
\alpha_2 > 1 + \frac{|\mathbf{k}_u|^2 - |\mathbf{k}_v|^2}{\epsilon - |\mathbf{k}_u|^2}\,,
\ee
while the state $(u,v)=(0,v)$ (corresponding to $\phi =\frac{\pi}{2}$) is locally stable on the absorbing set $\mathcal{A}$ provided that 
\be \label{die2}
\alpha_1 > 1 + \frac{|\mathbf{k}_v|^2 - |\mathbf{k}_u|^2}{\epsilon - |\mathbf{k}_v|^2}\,.
\ee

Although these conditions may appear to differ in form from those of the more general case \cite{van2018generic}, the difference is superficial, and due to the alternative derivation used in \cite{van2018generic}. Here we have used the explicit form of the saturable nonlinearity, which has allowed for an explicit construction of the conditions. Meanwhile, due to the general form of the nonlinearity used in \cite{van2018generic}, the conditions in that paper are expressed in terms of a local inverse function of the nonlinearity in a neighborhood of each amplitude death regime. If one takes the saturable nonlinearities, and inputs them into the conditions of \cite{van2018generic}, then one will obtain equivalent conditions to those we list here (after some changes in notation).

While the conditions we obtain provide local stability conditions for the relevant amplitude death regimes, we shall later demonstrate through numerical simulations that for many cases these amplitude death dynamics are globally stable for the relevant wavefunction. Note that the amplitude death of one wavefunction does not necessarily restrict the dynamics of the other wavefunction; indeed, the other wavefunction may exhibit a wide range of behaviors seen in scalar complex GL systems, as our simulations will later show.

\subsection{Asymptotic dynamics from a single scalar saturable complex GL equation}
While the system \eqref{sgl1}-\eqref{sgl2} is novel, note that a complex scalar GL equation with saturable nonlinearity has not been considered in the literature, with only NLS equations with saturable nonlinearity attracting attention thus far. Therefore, the scalar equation is of interest in its own right. However, note that obtaining results for such a complex scalar equation is not quite as simple as setting the XPM terms to zero (i.e. $\alpha_1 = \alpha_2 =0$), as the quantity $|u|^2 + |v|^2$ appears in the denominator of the system \eqref{sgl1}-\eqref{sgl2}, so one should start with the scalar equation and directly obtain the needed results. We sketch the obtained results here, for sake of completeness. To begin with, the scalar complex saturable GL equation reads
\be \label{single_sgl}
\frac{\pa u}{\pa t} = \epsilon u + \left( 1 + ia\right)\nabla^2 u -  \frac{\left( 1 - ib\right) |u|^2 u}{\left\lbrace 1+ \mu  |u|^{2n} \right\rbrace^{1/n} }\,.
\ee

To approximate the absorbing set, we again write 
\be 
u = e^{i\mathbf{k}\cdot\mathbf{x}}U(t)=\rho(t) e^{i \theta(t)}e^{i\mathbf{k}\cdot\mathbf{x}},
\ee
where 
\be 
\theta = \left(\epsilon b-(a+b) |\mathbf{k}|^2\right) t-b \log(\left|\rho\right|)
\ee
and the dynamics of $\rho$ are given by
\begin{equation}
\frac{\partial \rho}{\partial t} = \epsilon \rho-|\mathbf{k}|^2 \rho-\frac{\rho^3}{\left(1+\mu \rho^{2 n}\right)^{1/n}}\,.
\end{equation}
Define the intensity for the scalar equation \eqref{single_sgl} as $I=|u|^2$. Note $I(0)=0$, $I(|u|^2)>0$ for non-zero $|u|$, and $I$ is unbounded as $|u|$ grows. Further, 
\begin{equation}
\frac{1}{2}\frac{dI}{dt}=F(\rho)=\rho^2 \left(\epsilon-|\mathbf{k}|^2 -\frac{\rho^2}{\left(1+\mu \rho^{2 n}\right)^{1/n}}\right)\,.
\end{equation}
Now, $F(\rho) < 0$ if $\epsilon < |\mathbf{k}|^2+\frac{\rho^2}{((1+\mu \rho^{2 n})^{1/n}}$, and for asymptotically large $\rho$ this gives the condition $\epsilon<\mu^{-1/n}$, which is the first of the two possible parameter restrictions we obtained in the two-component case. Further, we have that $F(\rho) > 0$ for $\rho\in (0,\rho_0)$ and $F(\rho) < 0$ for $\rho \in (\rho_0, \infty)$, where 
\begin{equation}
\rho_0=\frac{\sqrt{\epsilon-|\mathbf{k}|^{2}}}{\left(1-\mu \left(\epsilon-|\mathbf{k}|^{2}\right)^{n}\right)^{1/\left(2 n\right)}} >0.
\end{equation}
Therefore, the norm $I$ will expand for $\rho < \rho_0$ and contract for $\rho > \rho_0$, making $\rho = \rho_0$ the absorbing set. These results are in line with the two-species case, only here we obtain a circle of radius $\rho_0$ in $\mathbb{C}$ as the absorbing set rather than a more complicated submanifold of $\mathbb{C}^2$. 

As $\rho_0 >0$, there will be no amplitude death for the single scalar complex GL equation, highlighting the fact that the mechanism behind amplitude death observed for the saturable system (and, for more general systems, in \cite{van2018generic}) does rely on the interaction of wavefunctions, and cannot be induced through only self-interaction terms.

\section{Plane wave solutions and their modulational instability}\label{modinst}
Having constructed an approximate absorbing set for the dynamics of \eqref{sgl1}-\eqref{sgl2}, we note that simulations generically show that solutions do not tend to globally uniform steady dynamics. This motivates us to study the modulational stability of plane wave solutions. As our simulations suggest, we are able show that for generic parameter combinations there will exist a family of unstable wavenumbers leading to modulational instability. As we shall show in Sec. \ref{numeric}, this gives the appearance of spatiotemporal chaos in the dynamics for many parameter regimes. We start out with plane wave solutions of constant amplitude, and perturb these linearly. The linear perturbations satisfy a linear partial differential equation system, which is then solved. The temporal decay or amplification of these perturbations will determine the stability or instability, respectively, of the plane waves. We shall show that there always exist perturbations of large wavenumbers (high-frequency perturbations) which result in loss of stability for the underlying plane wave solutions, and hence plane waves are generically modulationally unstable for the parameter regime we are concerned with.

\subsection{Construction of plane wave solutions}
We begin by looking for plane wave solutions
\begin{equation}\begin{aligned}\label{pwsol}
u(\mathbf{x},t) & = A e^{i\left(\mathbf{k}_u \cdot \mathbf{x} - \omega_u t \right)}, \\  
v(\mathbf{x},t) & = B e^{i\left(\mathbf{k}_v \cdot \mathbf{x} - \omega_v t \right)},
\end{aligned}\end{equation}
where $A,B,\omega_u,\omega_v \in \mathbb{R}$ and $\mathbf{k}_u,\mathbf{k}_v \in \mathbb{R}^n$. Substituting these solutions into \eqref{sgl1}-\eqref{sgl2} and separating the real and imaginary parts we find
\be
0 = \epsilon - |\mathbf{k}_u|^2 - \frac{A^2 + \alpha_1 B^2}{\left[1 + \mu \left( A^2 + B^2 \right)^n \right]^{\frac{1}{n}}}, 
\ee
\be
\omega_u  = a |\mathbf{k}_u|^2 - b \frac{A^2 + \alpha_1 B^2}{\left[1 + \mu \left( A^2 + B^2 \right)^n \right]^{\frac{1}{n}}},
\ee
\be 
0 = \epsilon - |\mathbf{k}_v|^2 - \frac{\alpha_2 A^2 + B^2}{\left[1 + \mu \left( A^2 + B^2 \right)^n \right]^{\frac{1}{n}}},
\ee
\be 
\omega_v  = a |\mathbf{k}_v|^2 - b \frac{\alpha_2 A^2 + B^2}{\left[1 + \mu \left( A^2 + B^2 \right)^n \right]^{\frac{1}{n}}},
\ee
from which we find
\begin{equation}
\omega_u = (a+b) |\mathbf{k}_u|^2 - b \epsilon, 
\end{equation}
\begin{equation}
\omega_v = (a+b) |\mathbf{k}_v|^2 - b \epsilon\,,
\end{equation}
\begin{equation}
A = \left[\frac{(1-\alpha_1 \alpha_2)^n}{\left\lbrace (1-\alpha_1) \epsilon - \left(|\mathbf{k}_u|^2 - \alpha_1 |
\mathbf{k}_v|^2\right)\right\rbrace^n} - \mu^n \left( 1 + \frac{(1-\alpha_2) \epsilon - \left(|\mathbf{k}_v|^2 - 
\alpha_2 |\mathbf{k}_u|^2\right)}{(1-\alpha_1) \epsilon - \left(|\mathbf{k}_u|^2  - \alpha_1 |\mathbf{k}_v|^2\right)}
\right)^n \right]^{-\frac{1}{2n}}\,,
\end{equation}
\begin{equation}\begin{aligned}
B & = \sqrt{\frac{(1-\alpha_2) \epsilon - \left(|\mathbf{k}_v|^2 - \alpha_2 |\mathbf{k}_u|^2\right)}{(1-\alpha_1) \epsilon - \left(|\mathbf{k}_u|^2 - \alpha_1 |\mathbf{k}_v|^2\right)}}\\
& \quad \times
\left[\frac{(1-\alpha_1 \alpha_2)^n}{\left\lbrace (1-\alpha_1) \epsilon - \left(|\mathbf{k}_u|^2 - \alpha_1 |
\mathbf{k}_v|^2\right)\right\rbrace^n} - \mu^n \left( 1 + \frac{(1-\alpha_2) \epsilon - \left(|\mathbf{k}_v|^2 - 
\alpha_2 |\mathbf{k}_u|^2\right)}{(1-\alpha_1) \epsilon - \left(|\mathbf{k}_u|^2  - \alpha_1 |\mathbf{k}_v|^2\right)}
\right)^n \right]^{-\frac{1}{2n}} \,.
\end{aligned}\end{equation}

The conditions for existence of plane wave solutions \eqref{pwsol} are therefore
\be 
\frac{(1-\alpha_1 \alpha_2)^n}{\left\lbrace (1-\alpha_1) \epsilon - \left(|\mathbf{k}_u|^2 - \alpha_1 |\mathbf{k}_v|^2\right)\right\rbrace^n} - \mu^n \left( 1 + \frac{ (1-\alpha_2) \epsilon - \left(|\mathbf{k}_v|^2 - \alpha_2 |\mathbf{k}_u|^2\right)}{(1-\alpha_1) \epsilon - \left(|\mathbf{k}_u|^2 - \alpha_1 |\mathbf{k}_v|^2\right)}\right)^n >0 
\ee
and 
\be 
\left\lbrace (1-\alpha_1) \epsilon - \left(|\mathbf{k}_u|^2 - \alpha_1 |\mathbf{k}_v|^2\right) \right\rbrace \left\lbrace (1-\alpha_2) \epsilon - \left(|\mathbf{k}_v|^2 - \alpha_2 |\mathbf{k}_u|^2\right)\right\rbrace > 0\,.
\ee
Under these conditions, the plane wave solutions will exist in the absorbing set $\mathcal{A}$. 

\subsection{Generic modulational instability of plane waves under small wavenumber perturbations}
In order to determine the stability or instability of the plane wave solutions \eqref{pwsol}, we follow the procedure described in \cite{zakeri2015modulational}. We first introduce perturbations to the plane wave solutions of the form
\begin{equation}\begin{aligned}\label{pwperts}
u(\mathbf{x},t) & = \left( A + \delta \hat{u}(\mathbf{x},t) \right) e^{i\left(\mathbf{k}_u \cdot \mathbf{x} - \omega_u t \right)},\\ 
v(\mathbf{x},t) & =  \left( B + \delta \hat{v}(\mathbf{x},t) \right) e^{i\left(\mathbf{k}_v \cdot \mathbf{x} - \omega_v t \right)},
\end{aligned}\end{equation}
where $\delta \ll 1$. Substituting the perturbed solutions \eqref{pwperts} into \eqref{sgl1}-\eqref{sgl2}, we find that the  $\mathcal{O}(\delta)$ perturbations are given by
\begin{subequations}\label{eq:hat_derivs}
\begin{align}
\frac{\partial \hat{u}}{\partial t} &= \epsilon \hat{u} + (1 + i a) \left( 2 i \mathbf{k}_u \cdot \nabla \hat{u} + \nabla^2 \hat{u} \right) - (1-i b) \left( \mathcal{F}_u \left(\hat{u} + \hat{u}^* \right) + \mathcal{F}_v \left( \hat{v} + \hat{v}^* \right) \right),\\
\frac{\partial \hat{v}}{\partial t} &= \epsilon \hat{v} + (1 + i a) \left( 2 i \mathbf{k}_v \cdot \nabla \hat{v} + \nabla^2 \hat{v} \right) - (1-i b) \left( \mathcal{G}_u \left(\hat{u} + \hat{u}^* \right) + \mathcal{G}_v \left( \hat{v} + \hat{v}^* \right) \right),
\end{align}
\end{subequations}
where $u^*$ is the complex conjugate, and the real-valued constants $\mathcal{F}_u$, $\mathcal{F}_v$, $\mathcal{G}_u$, and $\mathcal{G}_v$ are given by
\be
\mathcal{F}_u = \frac{A^2}{\left[1 + \mu \left(A^2 + B^2\right)^n \right]^{\frac{1}{n}}} \left( 1 - \frac{\mu (A^2 + \alpha_1 B^2)}{(A^2+B^2)^{1-n}\left(1 + \mu (A^2 + B^2)^n \right)} \right),
\ee
\be 
\mathcal{F}_v = \frac{AB}{\left[1 + \mu \left(A^2 + B^2\right)^n \right]^{\frac{1}{n}}} \left( \alpha_1 - \frac{\mu (A^2 + \alpha_1 B^2)}{(A^2+B^2)^{1-n}\left(1 + \mu (A^2 + B^2)^n \right)} \right),
\ee
\be 
\mathcal{G}_u = \frac{AB}{\left[1 + \mu \left(A^2 + B^2\right)^n \right]^{\frac{1}{n}}} \left( \alpha_2 - \frac{\mu (\alpha_2 A^2 + B^2)}{(A^2+B^2)^{1-n}\left(1 + \mu (A^2 + B^2)^n \right)} \right),
\ee
\be 
\mathcal{G}_v = \frac{B^2}{\left[1 + \mu \left(A^2 + B^2\right)^n \right]^{\frac{1}{n}}} \left( 1 - \frac{\mu (\alpha_2 A^2 + B^2)}{(A^2+B^2)^{1-n}\left(1 + \mu (A^2 + B^2)^n \right)} \right).
\ee

We take the ansatz
\begin{equation}\begin{aligned}\label{assume1}
\hat{u} & = A_1 e^{i (\mathbf{K}\cdot \mathbf{x} - \Omega t)} + A_2^* e^{-i (\mathbf{K}\cdot \mathbf{x} - \Omega^* t)},\\
\hat{v} & = B_1 e^{i (\mathbf{K}\cdot \mathbf{x} - \Omega t)} + B_2^* e^{-i (\mathbf{K}\cdot \mathbf{x} - \Omega^* t)},
\end{aligned}\end{equation}
which puts \eqref{eq:hat_derivs} into the form of the matrix system
\begin{equation}\label{eq:LS_pert}
(M- i\Omega I_{4\times 4})\begin{pmatrix}
A_1 \\
B_1 \\
A_2 \\
B_2
\end{pmatrix} 
= \begin{pmatrix}
M_{11} - i \Omega & M_{12} & M_{13} & M_{14} \\
M_{21} & M_{22} - i \Omega & M_{23} & M_{24} \\
M_{31} & M_{32} & M_{33} - i \Omega & M_{34} \\
M_{41} & M_{42} & M_{43} &  M_{44} - i \Omega
\end{pmatrix}
\begin{pmatrix}
A_1 \\
B_1 \\
A_2 \\
B_2
\end{pmatrix} = 
\begin{pmatrix}
0 \\
0 \\
0 \\
0
\end{pmatrix},
\end{equation}
where the elements of $M$ are given by
$$\begin{aligned}
M_{11} & = -\epsilon + (1 + i a) \left( 2 \mathbf{k}_u \cdot \mathbf{K} + |\mathbf{K}|^2 \right) + ( 1 - i b) \mathcal{F}_u,\\
M_{12} & = M_{14} =  (1 - i b) \mathcal{F}_v ,\\
M_{32} & = M_{34} = (1 + i b) \mathcal{F}_v,\\
M_{13} & = (1 - i b) \mathcal{F}_u ,\\
M_{31} & = (1 + i b) \mathcal{F}_u ,\\
M_{21} & = M_{23} = (1 - i b) \mathcal{G}_u , \\
M_{41} & = M_{43} = (1 + i b) \mathcal{G}_u , \\
M_{22} & = - \epsilon + (1 + i a) \left( 2 \mathbf{k}_v \cdot \mathbf{K} + |\mathbf{K}|^2 \right) + ( 1 - i b) \mathcal{G}_v,\\
M_{24} & =  (1-i b) \mathcal{G}_v , \\
M_{42} & = (1+i b) \mathcal{G}_v ,\\
M_{33} & = - \epsilon+ (1 - i a) \left( - 2 \mathbf{k}_u \cdot \mathbf{K} + |\mathbf{K}|^2 \right) + ( 1 + i b) \mathcal{F}_u,\\
M_{44} & = - \epsilon + (1 - i a) \left( - 2 \mathbf{k}_v \cdot \mathbf{K} + |\mathbf{K}|^2 \right) + ( 1 + i b) \mathcal{G}_v.
\end{aligned}
$$
In order to obtain a non-trivial solution for the amplitudes, we require the determinant of the matrix $M$ to be zero, giving a characteristic polynomial for $\Omega$: 
$$
\Omega^4 + \Lambda_3 \Omega^3 + \Lambda_2 \Omega^2 + \Lambda_1 \Omega + \Lambda_0 = 0,
$$
where the $\Lambda_j$'s are functions of the entries in $M$. We find that 
\begin{equation}
\Lambda_3 = i \left(M_{11}+M_{22}+M_{33}+M_{44}\right).
\end{equation} 
From the form of \eqref{assume1}, the plane waves \eqref{pwsol} will be unstable if $\text{Im}(\Omega) > 0$ for at least one of the four complex roots. 

Note that if $\text{Im}(\Lambda_3) < 0$, then at least one of the roots has positive imaginary part. To show this, assume we have a fourth order polynomial with roots $\Omega_j$ with $j =1,\dots,4$. Then, we can show (from Vieta's formula) that
\begin{equation}
\Lambda_3 = -\sum_{j=1}^4 \text{Re}(\Omega_j) - i \sum_{j=1}^4 \text{Im}(\Omega_j)\,.
\end{equation}
Therefore, if $\text{Im}(\Lambda_3) < 0$, then $\sum_{j=1}^4 \text{Im}(\Omega_j) > 0$, which means that at least one of the roots $\Omega_j$ must have positive imaginary part. Therefore, $\text{Im}(\Lambda_3) < 0$ is a sufficient condition for instability.

From the definition of $M$ we have that 
\begin{equation}
\Lambda_3 = i \left(M_{11}+M_{22}+M_{33}+M_{44}\right) = -4 a \left(\mathbf{k}_u + \mathbf{k}_v\right) \cdot \mathbf{K} + i\left(4 |\mathbf{K}|^2 + 2\left(\mathcal{F}_u + \mathcal{G}_v \right) - 4\epsilon\right),
\end{equation}
so the instability condition is given by 
\begin{equation}
\text{Im} (\Lambda_3) = 4 |\mathbf{K}|^2 + 2\left(\mathcal{F}_u + \mathcal{G}_v \right)  - 4\epsilon < 0,
\end{equation}
which is satisfied for any $\mathbf{K}$ such that
\begin{equation}\label{instab}
|\mathbf{K}|^2 < \epsilon -\frac{\mathcal{F}_u + \mathcal{G}_v}{2}.
\end{equation}
As we can choose $\mathbf{K}$ arbitrarily, we conclude that the system is always modulationally unstable on an unbounded spatial domain, for sufficiently small wavenumber perturbations. Meanwhile, when the spatial domain is finite, a plane wave may or may not be modulationally stable, depending on whether any of the admissible wavenumbers for that finite spatial domain satisfy the condition \eqref{instab}.

The result \eqref{instab} is consistent with earlier results for scalar complex GL equations which show instability of plane waves against small wavenumber perturbations \cite{chate1994spatiotemporal, bretherton1983intermittency, nozaki1983pattern, bartuccelli1990possibility}. This is also consistent with what is known of modulational instabilitites arising in coupled cubic-quintic complex GL systems \cite{porsezian2009modulational, alcaraz2010modulational}. Our derivation also supports the view that spatiotemporal chaos is ubiquitous in many parameter regimes when there is sufficient spatial variation in the initial solutions. On the other hand, when patterns emerge, such as banded features, they appear broad and on the same scale as the spatial domain, and may correspond to large wavenumbers. Such features often persist when they appear in numerical simulations.

\section{The singular limit $a+b=0$ and stationary states}\label{sec2e}
Singular limits are known to allow for integrable dynamics in otherwise non-integrable GL equations \cite{conte1993linearity}. In the limit where $a+b=0$, the nonlinearity and dispersion are in balance, and for the scalar complex GL equation this limit is known to permit exact solutions \cite{conte1993linearity} since this corresponds to a degenerate yet integrable limit. Placing $b=-a$ into \eqref{sgl1}-\eqref{sgl2}, we have
\be \label{ab1a}
\left( \frac{1-ia}{1+a^2}\right)\frac{\pa u}{\pa t} = \epsilon\left( \frac{1-ia}{1+a^2}\right) u + \nabla^2 u - \frac{\left(|u|^2 + \alpha_1 |v|^2\right)u}{\left\lbrace 1+ \mu \left( |u|^2 + |v|^2\right)^n\right\rbrace^{1/n} }\,,
\ee
\be \label{ab1b} 
\left( \frac{1-ia}{1+a^2}\right)\frac{\pa v}{\pa t} = \epsilon\left( \frac{1-ia}{1+a^2}\right) v + \nabla^2 v - \frac{\left(\alpha_2 |u|^2 + |v|^2\right)v}{\left\lbrace 1+ \mu \left( |u|^2 + |v|^2\right)^n\right\rbrace^{1/n} }\,.
\ee
In the language of \cite{conte1993linearity}, this defines a system of \textit{real} GL equations since the coefficients of dispersion and the nonlinear terms are real. However, as the coefficient of the time derivatives has a non-zero real part, the system is not reducible to a type of NLS system (for which there are some results on exactly solvable cases), as there will be amplitude gain or decay that precludes a solitary wave which maintains its form as it propagates. In the scalar case, there have been efforts to determine integrability and to classify families of exact solutions to specific forms of real GL equations \cite{keefe1986integrability,gagnon1989exact,gagnon1988lie,gagnon1989lie,gagnon1989lie2}. However, since we have a system with non-polynomial nonlinearity and temporal amplitude gain or loss, we are not aware of any efforts to obtain exact solutions for such problems. It may be possible to systematically study solutions in the $a+b=0$ case for a related problem where there is no amplitude gain or loss, however, and we leave this as future work.

Following the approach of \cite{newton1986instabilities}, where spatially periodic solutions were considered for the complex cubic GL equation, one may assume a separation of space and time variables, with the time entering into a complex exponential. Assuming a stationary solution $u(\mathbf{x},t)=e^{-i\omega_R t}R(\mathbf{x})$, $v(\mathbf{x},t)=e^{-i\omega_S t}S(\mathbf{x})$ for real $R$ and $S$, we obtain from \eqref{ab1a}-\eqref{ab1b} the system
\be 
-\frac{i\omega_R + a \omega_R}{1+a^2}R =  \epsilon \left( \frac{1-ia}{1+a^2} \right)R + \nabla^2 R   - \frac{R^2 + \alpha_1 S^2}{\left\lbrace 1+ \mu \left( R^2 + S^2\right)^n\right\rbrace^{1/n} }  R = 0\,,
\ee
\be 
-\frac{i\omega_S + a \omega_S}{1+a^2}S =  \epsilon \left( \frac{1-ia}{1+a^2} \right)S + \nabla^2 S - \frac{\alpha_2 R^2 + S^2}{\left\lbrace 1+ \mu \left( R^2 + S^2\right)^n\right\rbrace^{1/n} }  S = 0\,.
\ee
Choosing
\be 
\omega_R = \omega_S = a\epsilon 
\ee
to remove the imaginary parts, and hence removing any secular growth, we see that the proper form of the stationary solutions are
\be 
u(\mathbf{x},t)=e^{-ia \epsilon t}R(\mathbf{x}) \quad \text{and} \quad v(\mathbf{x},t)=e^{-ia \epsilon t}S(\mathbf{x})\,,
\ee
where $R$ and $S$ satisfy the stationary PDE system
\be \label{ab3a}
 \nabla^2 R + \left\lbrace \epsilon  - \frac{R^2 + \alpha_1 S^2}{\left\lbrace 1+ \mu \left( R^2 + S^2\right)^n\right\rbrace^{1/n} } \right\rbrace R = 0\,,
\ee
\be \label{ab3b} 
 \nabla^2 S + \left\lbrace \epsilon - \frac{\alpha_2 R^2 + S^2}{\left\lbrace 1+ \mu \left( R^2 + S^2\right)^n\right\rbrace^{1/n} } \right\rbrace S = 0\,,
\ee
The system \eqref{ab3a}-\eqref{ab3b} is an NLS system with saturable nonlinearity. Existence of solutions to NLS systems of the form \eqref{ab3a}-\eqref{ab3b}, under certain parameter restrictions, were considered in \cite{de2013weakly}.

\subsection{Conditions for spatially oscillating solutions in one spatial dimension}
In the case where the spatial domain is one dimensional, i.e., $\mathbb{R}$, then bounded spatially periodic or non-periodic oscillatory stationary solutions are possible. Sufficient conditions for the existence of bounded oscillatory solutions in one spatial dimension are found separately to be
\be 
\epsilon > \max_{p,q >0} \frac{p + \alpha_1 q}{\left\lbrace 1+ \mu \left( p + q\right)^n\right\rbrace^{1/n} } = \frac{\max\left\lbrace 1, \alpha_1 \right\rbrace}{\mu^{1/n}}
\ee
and
\be 
\epsilon > \max_{p,q >0} \frac{\alpha_2 p + q}{\left\lbrace 1+ \mu \left( p + q\right)^n\right\rbrace^{1/n} } = \frac{\max\left\lbrace 1, \alpha_2 \right\rbrace}{\mu^{1/n}}\,.
\ee
Hence, we take 
\be \label{ineq1}
\epsilon >  \frac{\max\left\lbrace 1, \alpha_1 , \alpha_2 \right\rbrace}{\mu^{1/n}} 
\ee
as a sufficient condition for oscillations in space. As we shall demonstrate later, there are parameter regimes for which spatially periodic solutions exist outside of this bound, so this condition is sufficient but not necessary. 

Since each of $R$ and $S$ may oscillate with a distinct spatial period when the coupling is removed, quasi-periodic solutions may be observed (with two periods), while some parameter regimes will permit periodic solutions.
When these conditions do not hold, bounded oscillatory solutions are still possible in some parameter regimes, as the amplitudes of $R$ and $S$ are bounded, and hence the maximum is not attained. However, when these conditions do not hold, there may be additional conditions needed to avoid destructive resonances. Standing solitary waves may also be possible from the system \eqref{ab3a}-\eqref{ab3b}, but obtaining conditions for their existence and then constructing such solutions would likely be highly involved, and may be considered elsewhere.

\subsection{Non-existence of integrable limits permitting closed-form solutions when $\alpha_1 \neq \alpha_2$}
Finding exact solutions to \eqref{ab3a} - \eqref{ab3b} is actually quite difficult, relative to doing the same form more standard NLS systems. In order to explicitly construct spatially periodic solutions, we attempted the approach of \cite{chow2006exact}. If one assumes that solutions to \eqref{ab3a} - \eqref{ab3b} take the form $R(\mathbf{x}) = \hat{R}(x) = A \text{sn}(Xx,\kappa)$ and $S(\mathbf{x}) = \hat{S}(x) = A \text{cn}(Xx,\kappa)$, where sn and cn denote Jacobi elliptic sine and cosine functions, respectively, and $A, X>0$, $\kappa \in [0,1]$ are constants to be determined, then we find that the resulting algebraic system for these constants gives $\kappa =0$. The reason is that in \cite{chow2006exact} the saturable parameter $\mu$ was permitted to be different between the two wavefunctions. In the case where these parameters were taken to be the same, the XPM parameters also were required to be equal, and to unity. In such a limit, $R^2 + S^2$ is constant, and the dynamics reduce to that of the standard cubic complex GL system with equal self- and cross-phase modulation parameters. Therefore, the asymmetry in the XPM parameters destroys the particular structure which permits solutions in \cite{chow2006exact}. 

While there are no (non-trivial, $\kappa =0$ corresponds to plane waves studied in the previous section) explicit closed form solutions given in terms of Jacobi elliptic functions, we may still consider the situation where $R^2 + S^2 =A^2$, which is a generalization of such solutions. Then, the solution pair should satisfy
\be \label{newode}
{R'}^2 + RR'' + {S'}^2 + SS'' = 0\,,
\ee
where prime denotes differentiation with respect to $x$. Using $R^2 + S^2 =A^2$, note that we may write \eqref{ab3a} - \eqref{ab3b} in the form
\be 
R'' + + \left\lbrace \epsilon  - \frac{\alpha_1 A^2}{\left\lbrace 1+ \mu A^{2n}\right\rbrace^{1/n} } \right\rbrace R -\frac{(1-\alpha_1)}{\left\lbrace 1+ \mu A^{2n}\right\rbrace^{1/n} } R^3 = 0\,,
\ee
\be 
S'' + + \left\lbrace \epsilon  - \frac{\alpha_2 A^2}{\left\lbrace 1+ \mu A^{2n}\right\rbrace^{1/n} } \right\rbrace S -\frac{(1-\alpha_2)}{\left\lbrace 1+ \mu A^{2n}\right\rbrace^{1/n} } S^3 = 0\,.
\ee
After several manipulations, we find
\be 
{R'}^2 = C_1 - \beta_1 R^2 + \frac{\gamma_1}{2}R^4 \,,
\ee
\be 
RR'' = -\beta_1 R^2 + \gamma_1 R^4\,,
\ee
\be 
{S'}^2 = C_2 - \beta_2 S^2 - \frac{\gamma_2}{2}S^4 = C_2 - \beta_2 (A^2 - R^2) + \frac{\gamma_2}{2}(A^2 - R^2)^2\,,
\ee
\be 
SS'' = - \beta_2 S^2 + \gamma_2 S^4 = - \beta_2 (A^2 - R^2) + \gamma_2 (A^2 - R^2)^2\,,
\ee
where
\be 
\beta_\ell = \epsilon - \frac{\alpha_\ell A^2}{\left\lbrace 1+ \mu A^{2n}\right\rbrace^{1/n} }\,, \quad 
\gamma_\ell = \frac{(1-\alpha_\ell) A^2}{\left\lbrace 1+ \mu A^{2n}\right\rbrace^{1/n} }\,,
\ee
and $C_\ell$ are integration constants. Placing these into \eqref{newode}, we find that
\be 
C_1 + C_2 - 2\beta_2 A^2 + \frac{3}{2}\gamma_2 A^4 + \left( 2(\beta_2 - \beta_1) - 3\gamma_2 A^2 \right) R^2 + \frac{3}{2}(\gamma_1 + \gamma_2) R^4 = 0\,. 
\ee
By linear independence of the different powers of $R$, the coefficients of $O(1)$, $O(R^2)$, $O(R^4)$ must all be zero. The $O(1)$ coefficient may be set to zero through an appropriate choice of integration constants. However, setting the $O(R^4)$ coefficient to zero, we have $\gamma_1 + \gamma_2 = 0$, which in turn implies $\alpha_1 + \alpha_2 = 2$, which is fairly restrictive of the XPM parameters. Setting the $O(R^2)$ coefficient to zero, we find the additional parameter restriction $2\alpha_1 + \alpha_2 = 3$. Combining these two restrictions on the XPM parameters, we have that $\alpha_1 = \alpha_2 = 1$ is a necessary condition for the existence of a solution pair such that $R^2 + S^2$ is constant. 

What we have demonstrated is that there are generically no solution pairs such that the quantity $R^2 + S^2$ is a constant, except for the case where all nonlinearity parameters are identical ($\alpha_1 = \alpha_2 =1$). This rules out solution pairs such as those considered in \cite{chow2006exact}.

\subsection{First integral for in-phase solutions}
Despite the difficulties in obtaining exact closed-form solutions in regimes different from $\alpha_1 = \alpha_2 =1$, there is a natural parameter regime where a first integral may be obtained. To this end, we consider the case where $S(x) = \sigma R(x)$ for some constant $\sigma \in \mathbb{R}$. Note that this case will satisfy $R^2 + S^2 = (1+\sigma^2)R^2$, which is not constant. However, this simplification will grant a degree of analytical tractability. Note that for such a case, the solutions $R$ and $S$ will remain in phase with one another.

For $\sigma \neq 0$, we have that \eqref{ab3a}-\eqref{ab3b} become
\be \label{ab4a}
R'' + \left\lbrace \epsilon  - \frac{(1+ \alpha_1 \sigma^2) R^2}{\left\lbrace 1+ \mu \left( 1+\sigma^2\right)^n R^{2n}\right\rbrace^{1/n} } \right\rbrace R = 0\,,
\ee
\be \label{ab4b} 
R'' + \left\lbrace \epsilon - \frac{(\alpha_2 + \sigma^2) R^2}{\left\lbrace 1+ \mu \left( 1+\sigma^2\right)^n R^{2n}\right\rbrace^{1/n} } \right\rbrace R = 0\,.
\ee
Observe that \eqref{ab4a}-\eqref{ab4b} are consistent only if
\be 
1 + \alpha_1 \sigma^2 = \alpha_2 + \sigma^2 \,.
\ee
Therefore, we have that
\be \label{sigma}
\sigma = \begin{cases}
\pm \sqrt{\frac{1-\alpha_2}{1-\alpha_1}}, & \text{if} ~~ \alpha_1 \neq 1 ~~ \text{and} ~~ \text{sgn}(1-\alpha_1) = \text{sgn}(1-\alpha_2),\\
\pm 1, & \text{if} ~~ \alpha_2 = \alpha_1 = 1\,.
\end{cases}
\ee
If the sign of $1-\alpha_1$ and $1-\alpha_2$ are not the same, or if $\alpha_2 \neq \alpha_1 = 1$, then the equations \eqref{ab4a}-\eqref{ab4b} are never consistent and hence there is no such $\sigma$ such that $S=\sigma R$. 

Assuming we are in a case where \eqref{sigma} holds, and hence the equations \eqref{ab4a}-\eqref{ab4b} are consistent, we may obtain a first integral of \eqref{ab4a}, via integration of
\be
2R'R'' + 2\left\lbrace \epsilon  - \frac{(1+ \alpha_1 \sigma^2) R^2}{\left\lbrace 1+ \mu \left( 1+\sigma^2\right)^n R^{2n}\right\rbrace^{1/n} } \right\rbrace  RR' = 0
\ee
over the spatial domain. Recalling that 
\be 
\int \frac{qdq}{(1+ q^n)^{1/n}} = \frac{q^2}{2} ~~ _2F_1\left( \frac{1}{n}, \frac{2}{n}; 1 + \frac{2}{n} ; - q^n \right)\,,
\ee
where $_2F_1$ denotes the relevant hypergeometric function, we find after several algebraic manipulations that
\be \label{first}
{R'}^2 + \epsilon R^2 - \frac{1+\alpha_1 \sigma^2}{2} R^4 ~~ _2F_1\left( \frac{1}{n}, \frac{2}{n}; 1 + \frac{2}{n} ; - \mu (1+ \sigma^2)^n R^{2n} \right) = \mathcal{C}\,,
\ee
where $\mathcal{C}$ is an integration constant. 

Equation \eqref{first} defines a first integral for $R$. Even for $n=1$, there will not exist a second integral, and hence we may not solve for $R$ via quadrature. Still, this $S= \sigma R$ case does provide an interesting limit where we reduce the dimension of the nonlinear system to be solved. From the first integral \eqref{first}, one may obtain level sets for fixed values of $\mathcal{C}$. If these level sets are closed in the $R - R'$ plane, then there exists an oscillatory solution for $R$ (and, hence, for $S$).

From the form of \eqref{first}, we see that the condition \eqref{ineq1} will influence the emergent dynamics of this reduced system. Indeed, if $\epsilon$ is not large enough, then the dominant balance will be between ${R'}^2$ and the nonlinear term involving the hypergeometric function. In this case, we will have that ${R'}^2 >0$ for all $x>x_0$ for some finite $x_0$, precluding an oscillatory solution. This would correspond to a level set in the $R - R'$ plane which is not closed, but rather is not contained within any bounded set near the origin. Such a level set would then correspond to a divergent trajectory in phase space.

\subsection{Behavior of stationary solutions which are oscillatory in space}
\begin{figure}
\centering
\vspace{-1in}
\includegraphics[width=0.3\textwidth]{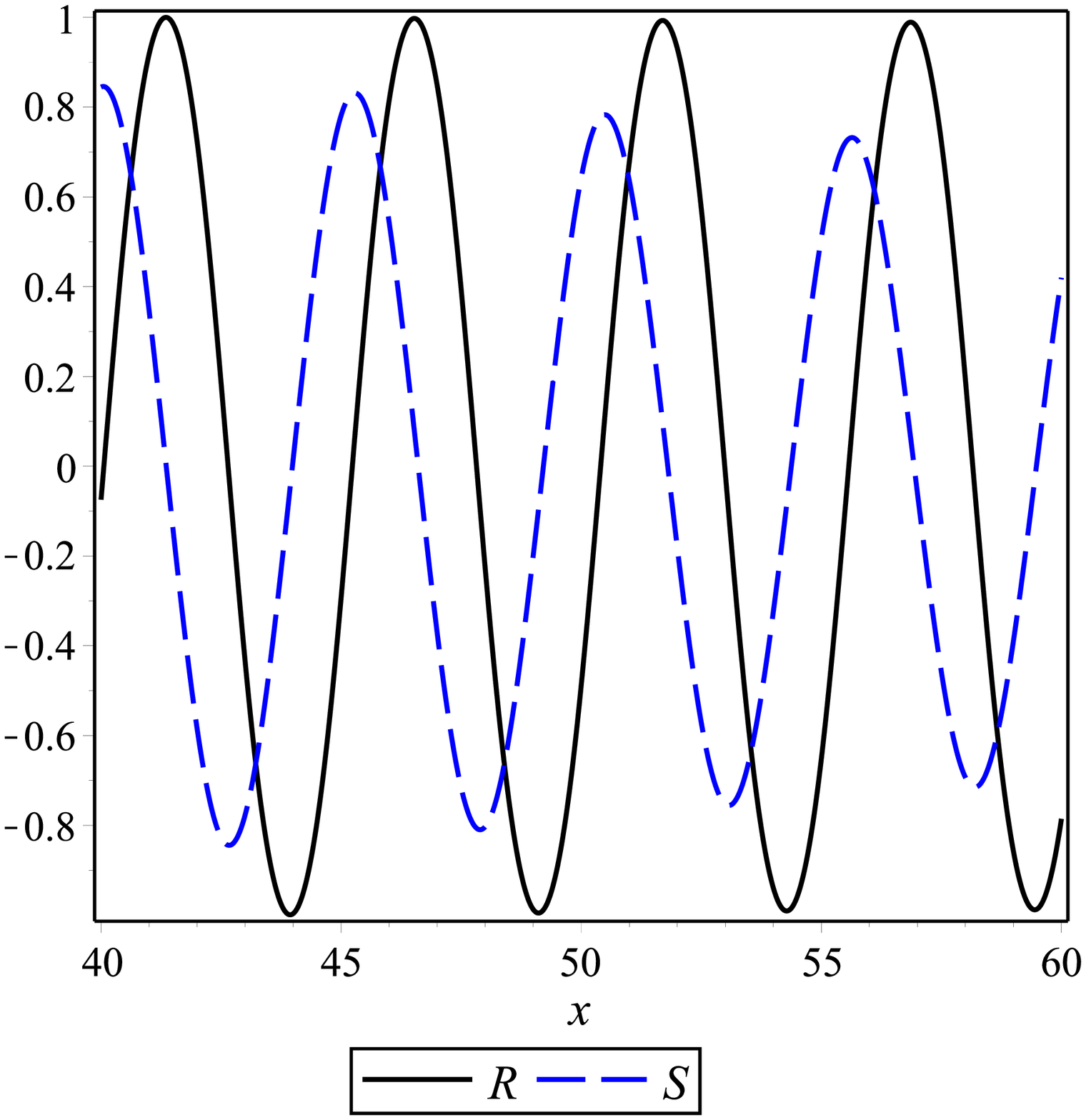}
\includegraphics[width=0.3\textwidth]{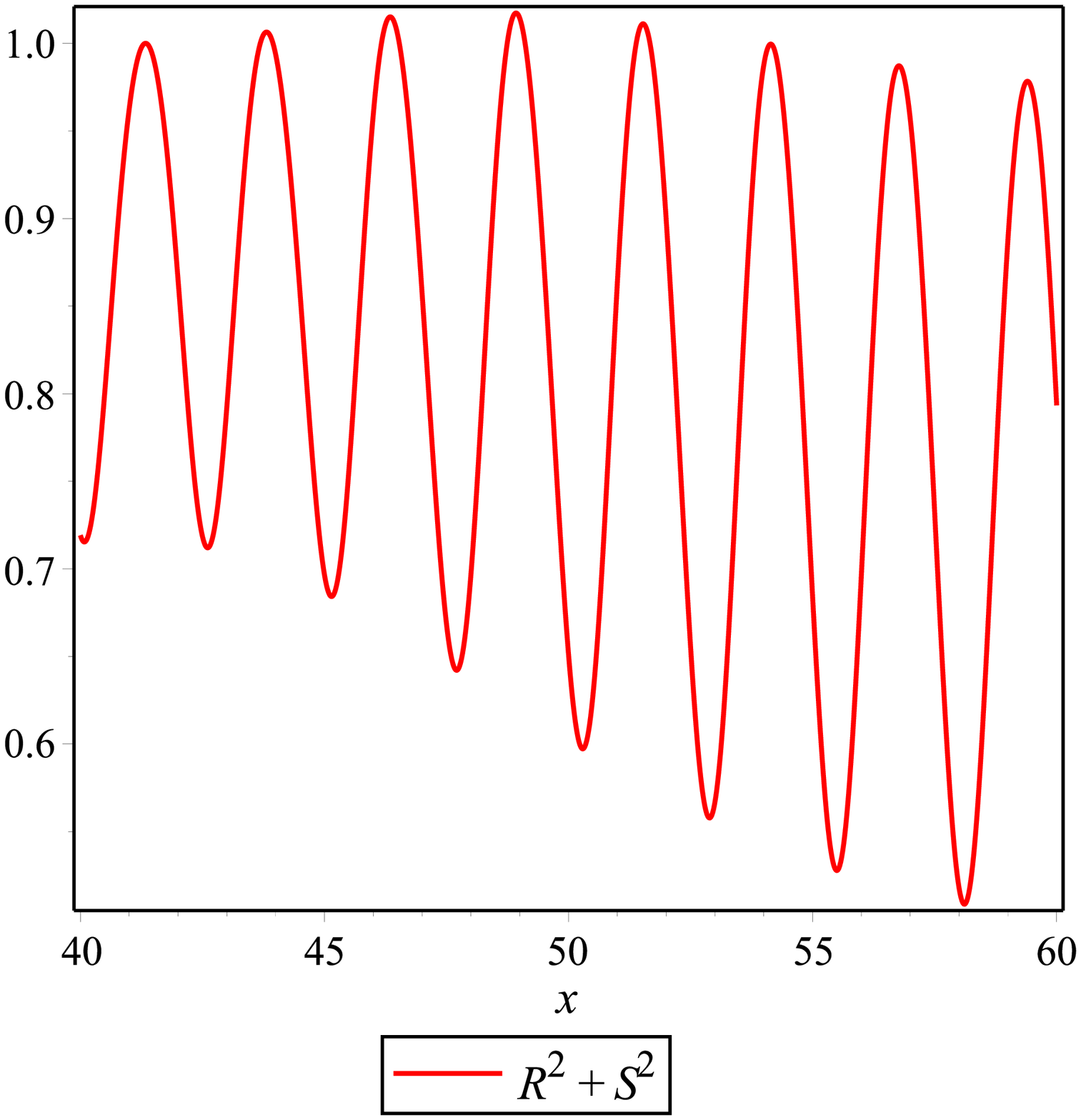}
\vspace{-0.2in}
$$
\text{(a)} \qquad\qquad\qquad\qquad\qquad\qquad \text{(b)}
$$
\vspace{-0.2in}
\includegraphics[width=0.3\textwidth]{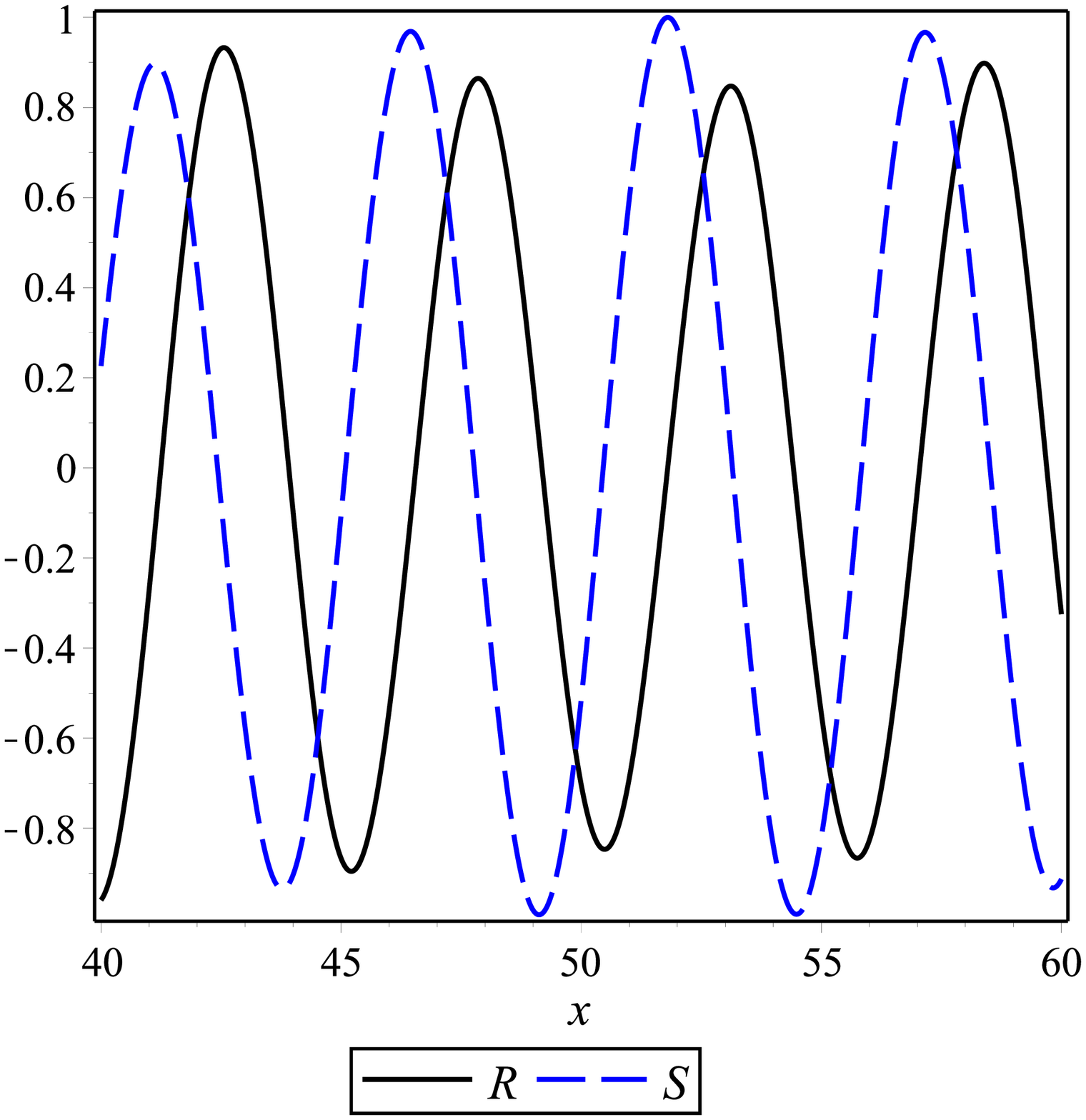}
\includegraphics[width=0.3\textwidth]{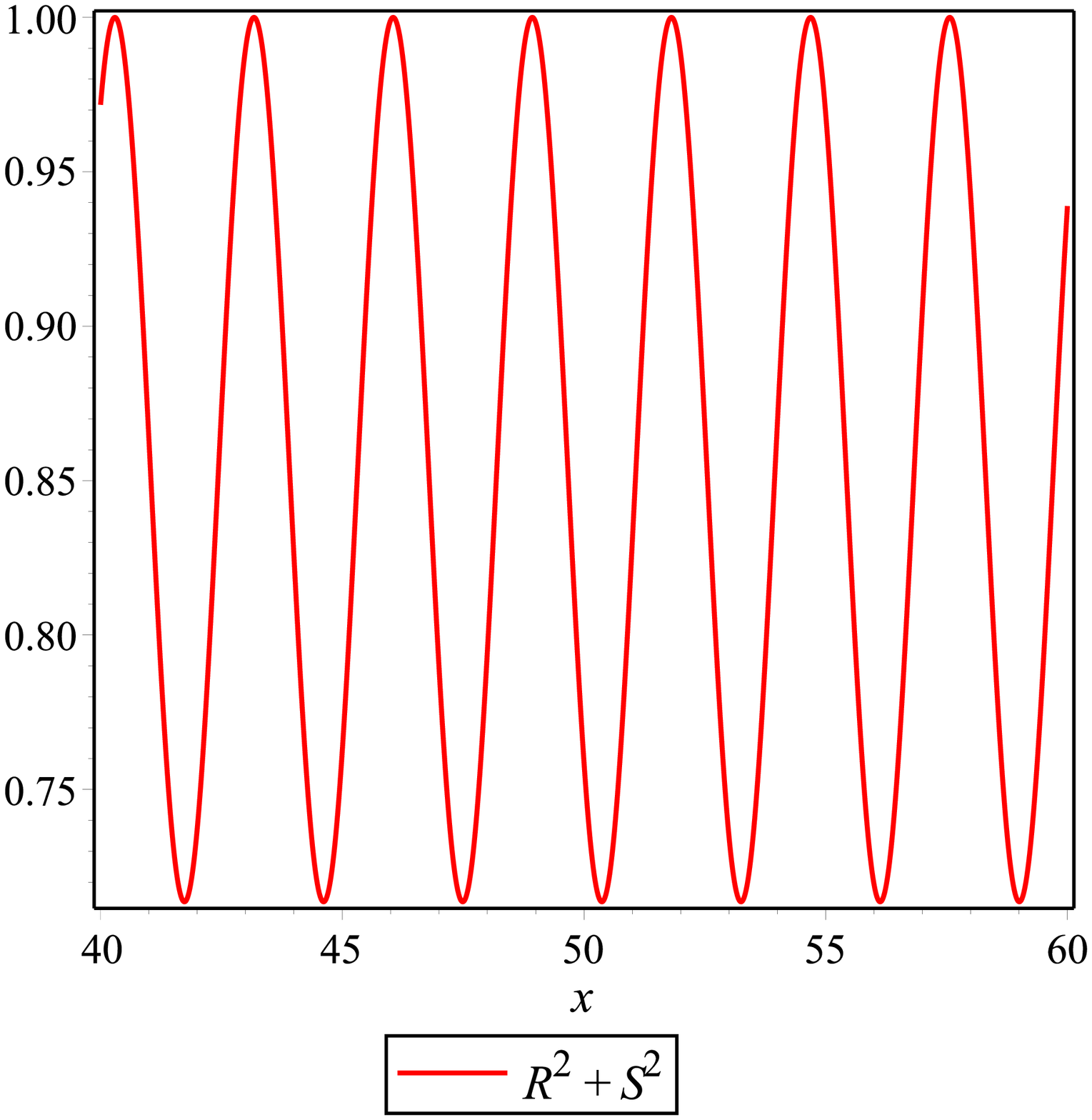}
$$
\text{(c)} \qquad\qquad\qquad\qquad\qquad\qquad \text{(d)}
$$
\vspace{-0.2in}
\includegraphics[width=0.3\textwidth]{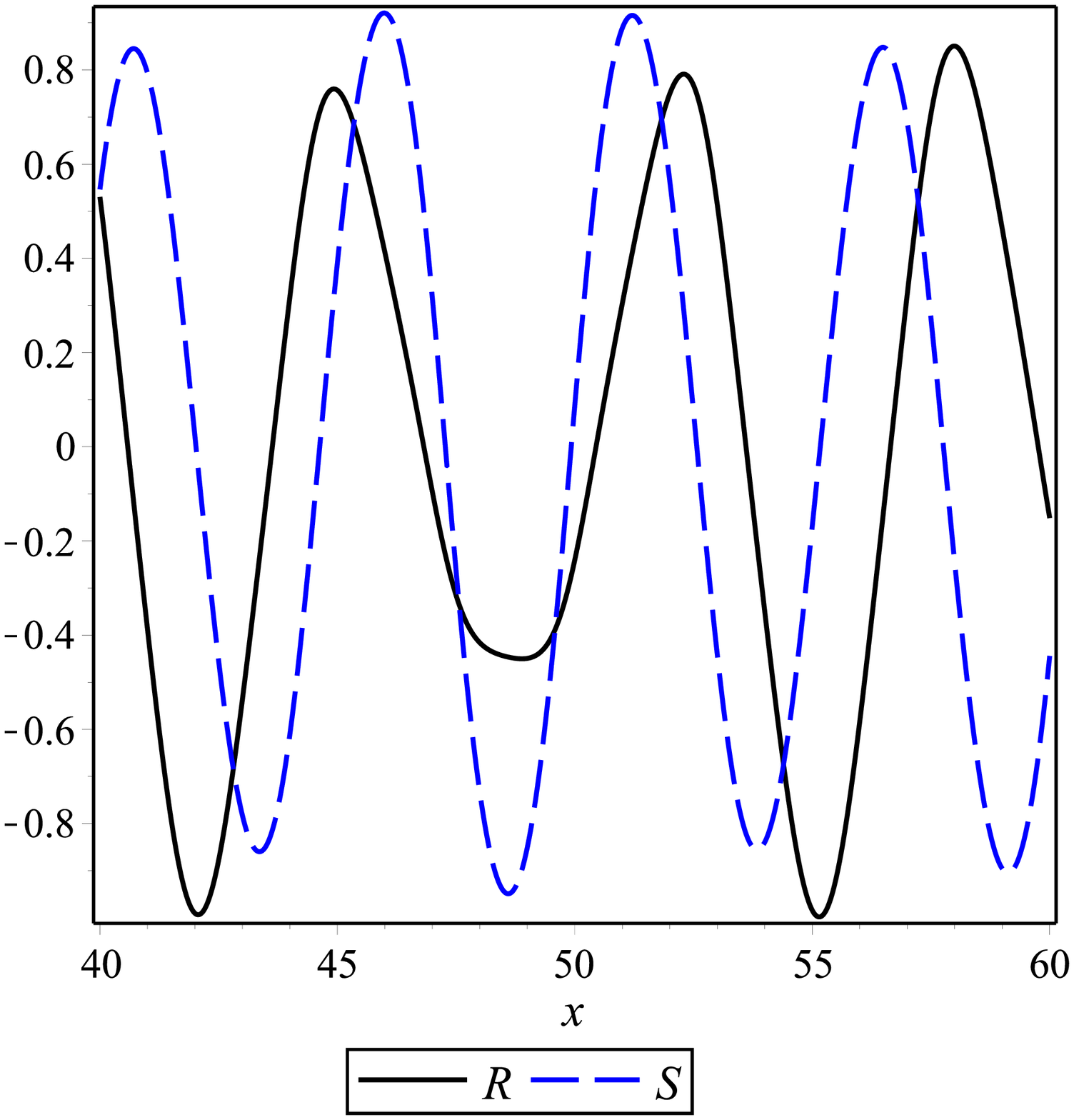}
\includegraphics[width=0.3\textwidth]{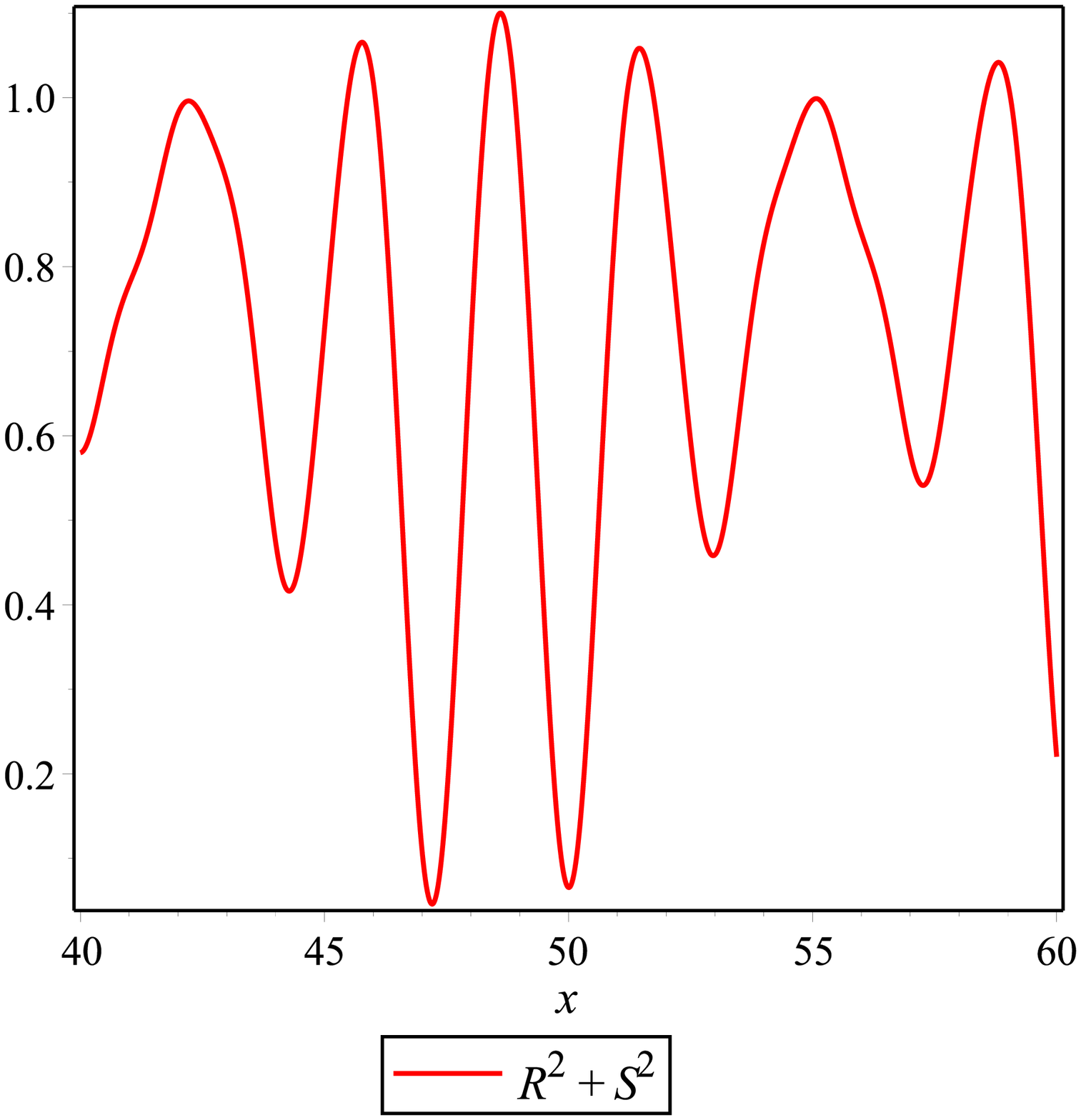}
$$
\text{(e)} \qquad\qquad\qquad\qquad\qquad\qquad \text{(f)}
$$
\vspace{-0.2in}
\includegraphics[width=0.3\textwidth]{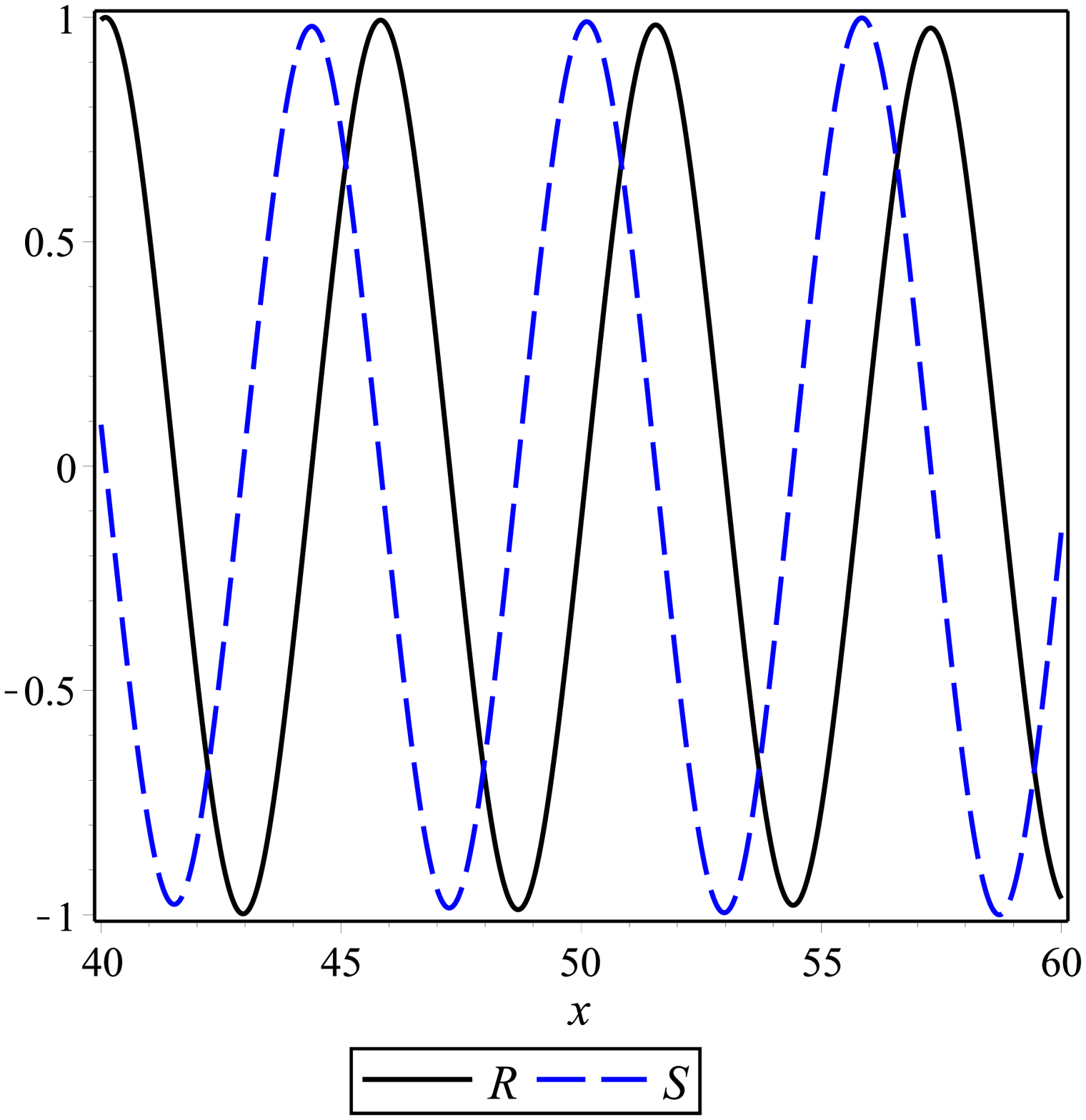}
\includegraphics[width=0.3\textwidth]{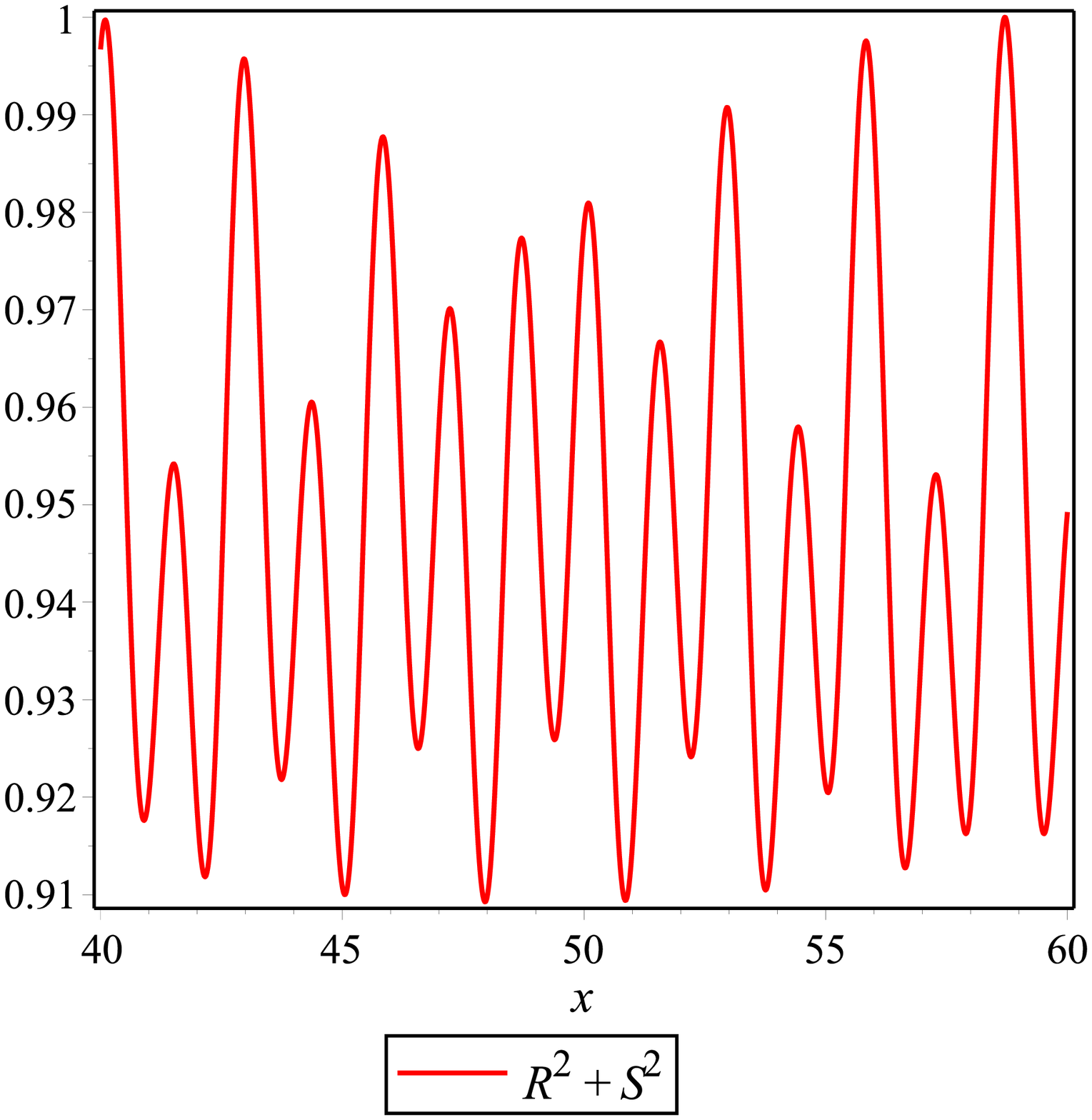}
$$
\text{(g)} \qquad\qquad\qquad\qquad\qquad\qquad \text{(h)}
$$
\vspace{-0.2in}
\caption{Plot of 1D stationary solutions $R$ and $S$ (left) and solution intensity $R^2 + S^2$ (right) governed by equations \eqref{ab3a}-\eqref{ab3b} for (a,b) $\alpha_1 = 0.1$, $\alpha_2 =1$, (c,d) $\alpha_1 = \alpha_2 =1$, (e,f) $\alpha_1 = 3$, $\alpha_2 =1$, (g,h) $\alpha_1 = \alpha_2 =2$. We set $\epsilon =2$, $\mu = 0.5$ and $n=1$, as well as $R(0)=1$, $R'(0)=0$, $S(0)=0$, $S'(0) =-1$ to obtain solutions which are out of spatial phase.}\label{figB}
\end{figure}

\begin{figure}
\centering
\vspace{-1in}
\includegraphics[width=0.3\textwidth]{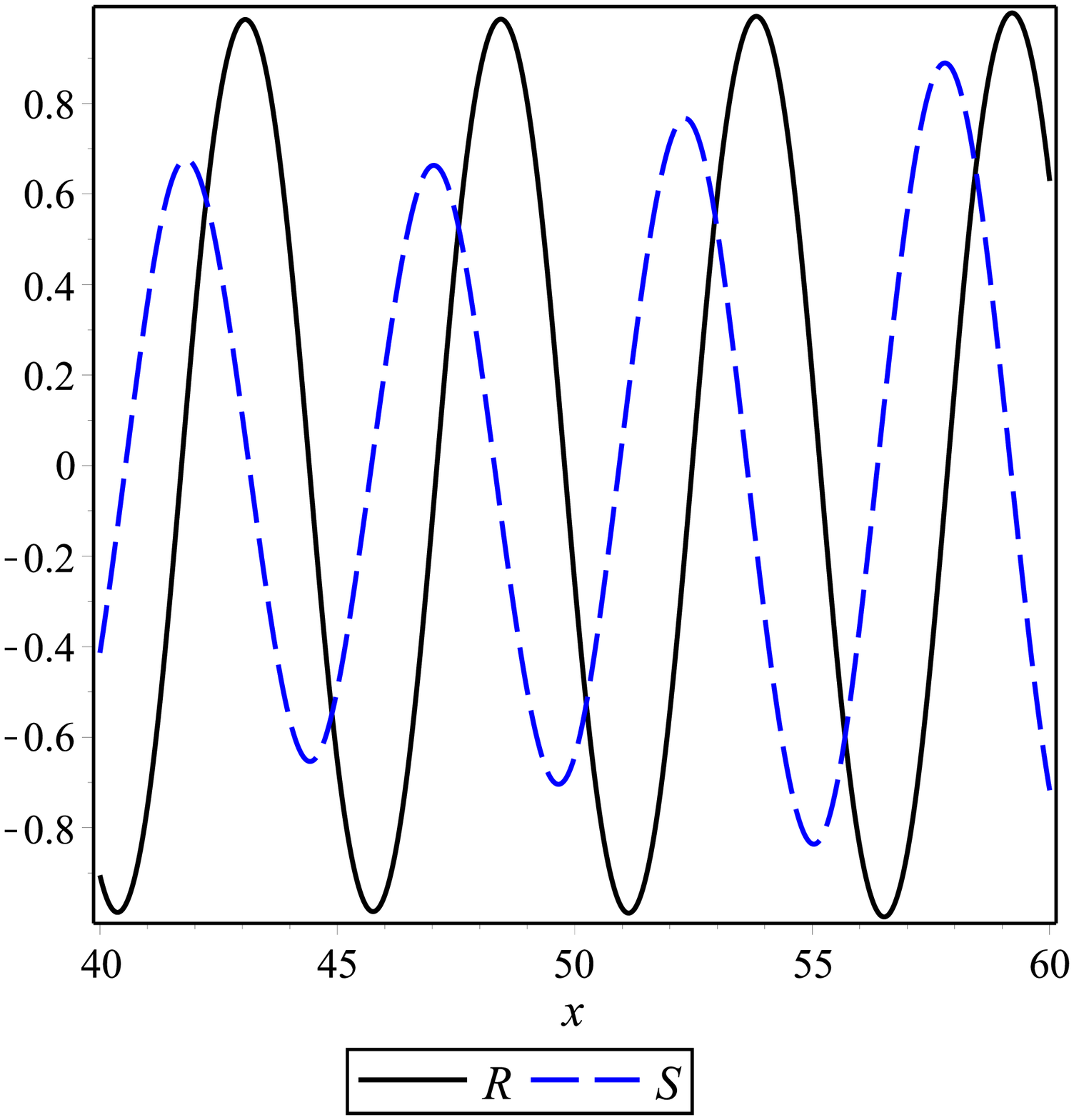}
\includegraphics[width=0.3\textwidth]{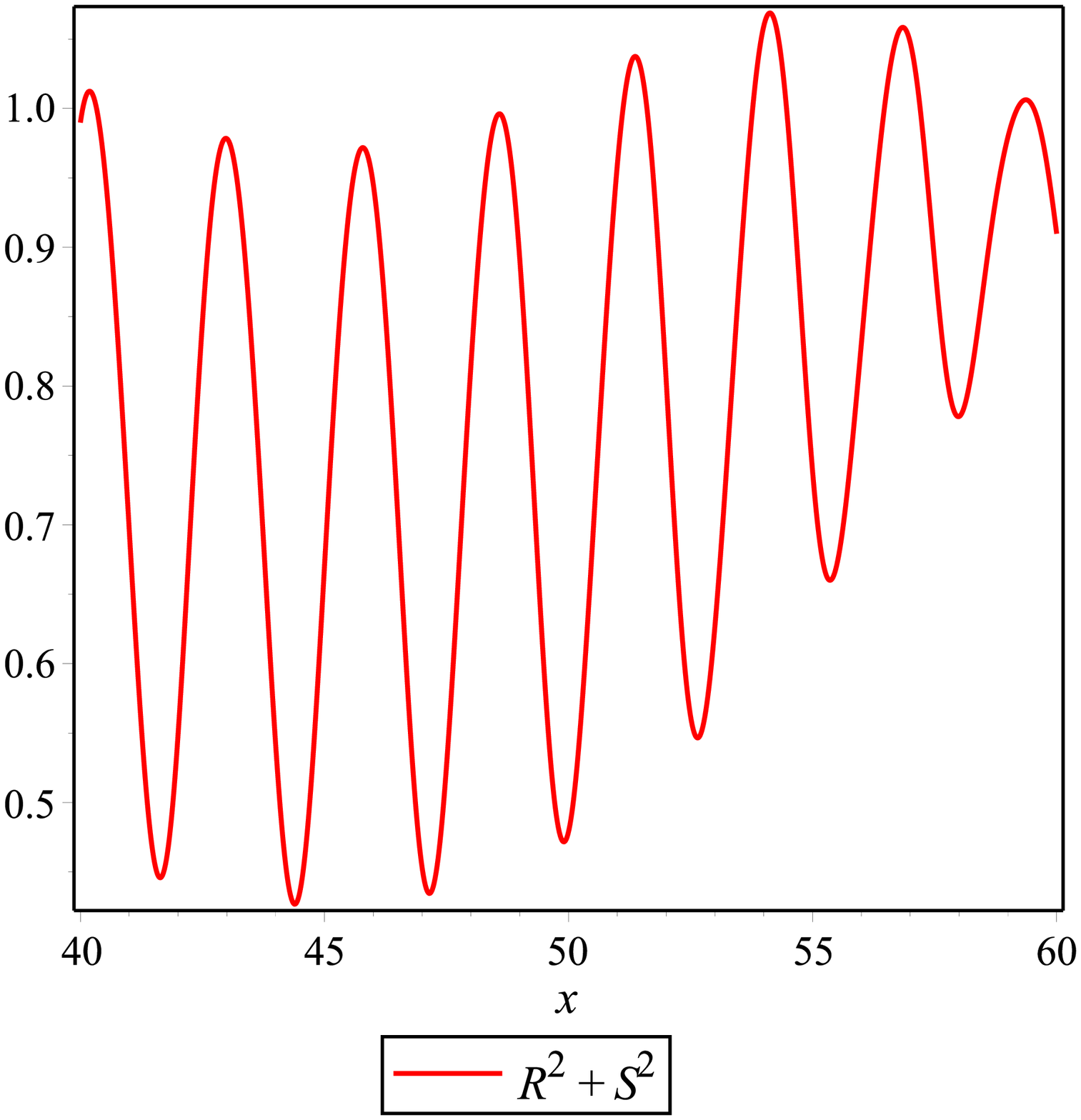}
\vspace{-0.2in}
$$
\text{(a)} \qquad\qquad\qquad\qquad\qquad\qquad \text{(b)}
$$
\vspace{-0.2in}
\includegraphics[width=0.3\textwidth]{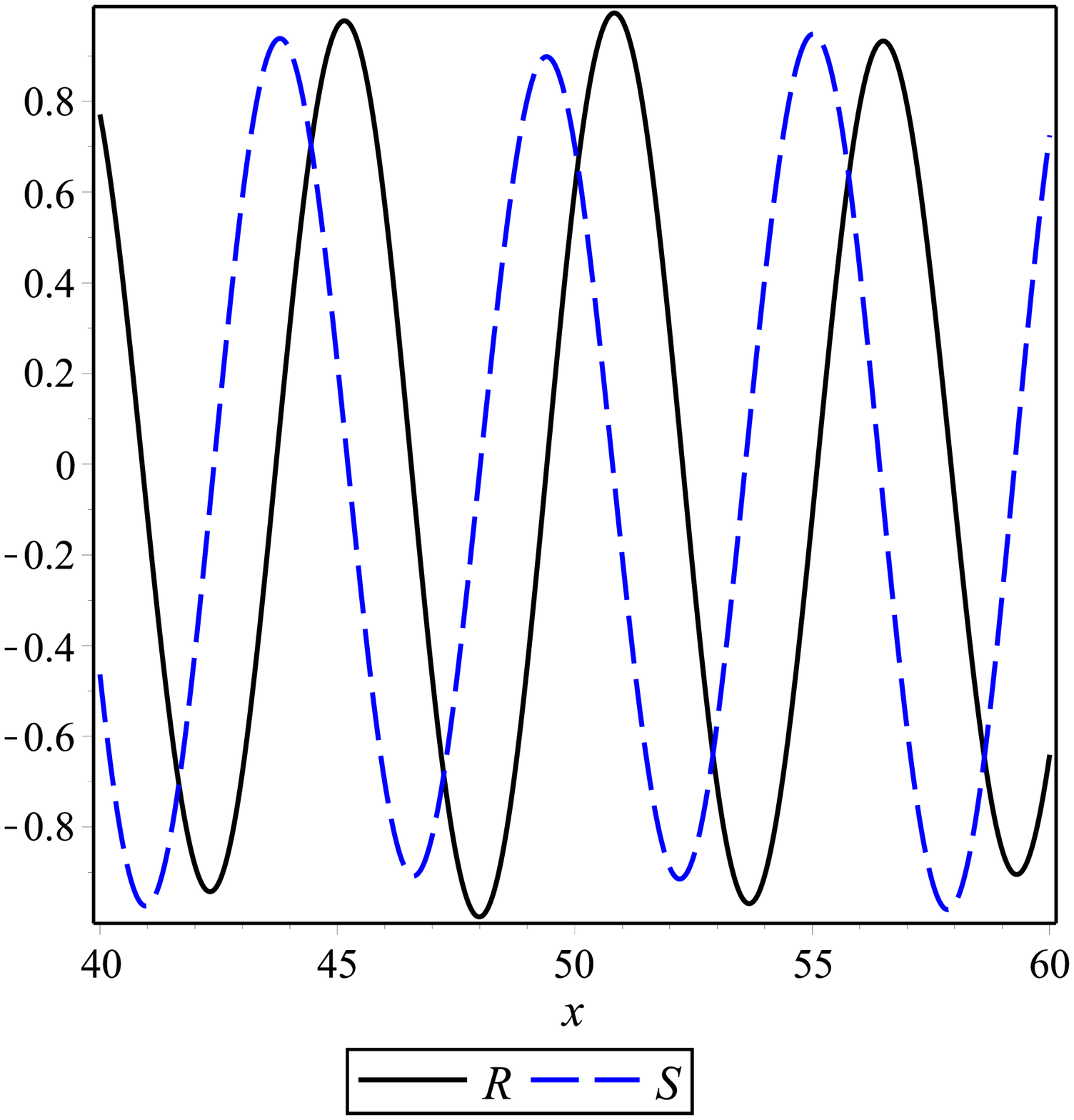}
\includegraphics[width=0.3\textwidth]{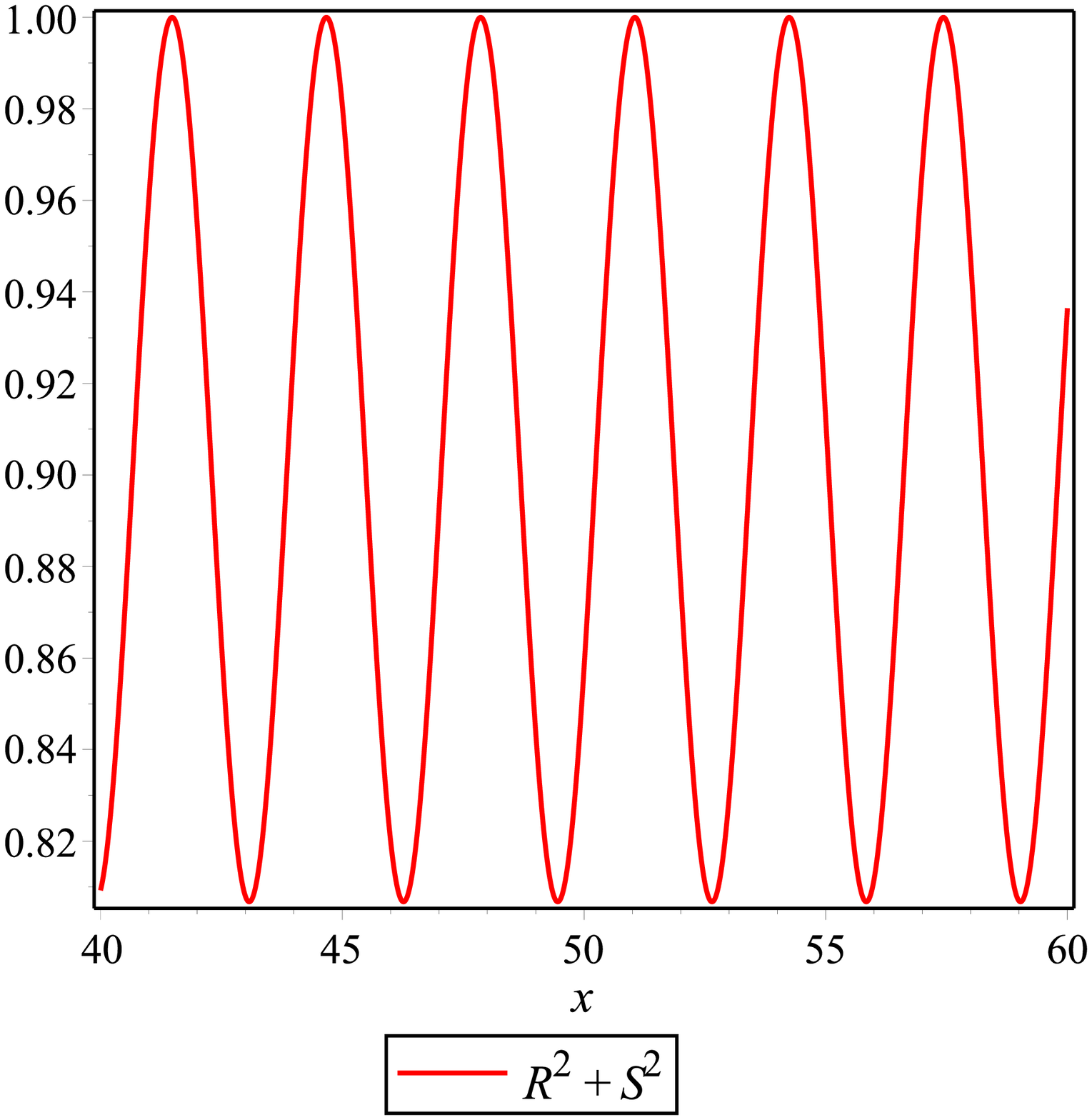}
$$
\text{(c)} \qquad\qquad\qquad\qquad\qquad\qquad \text{(d)}
$$
\vspace{-0.2in}
\includegraphics[width=0.3\textwidth]{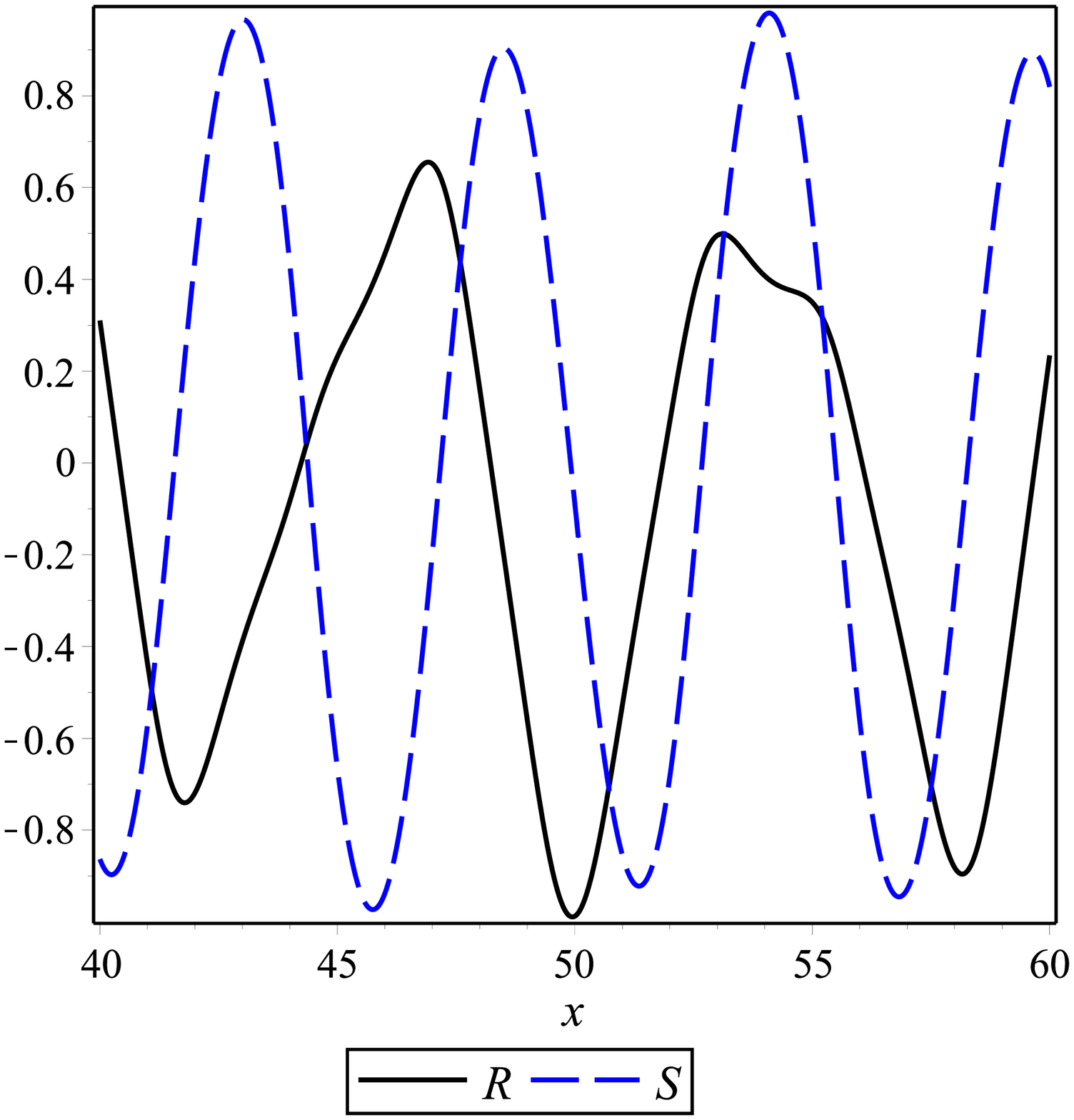}
\includegraphics[width=0.3\textwidth]{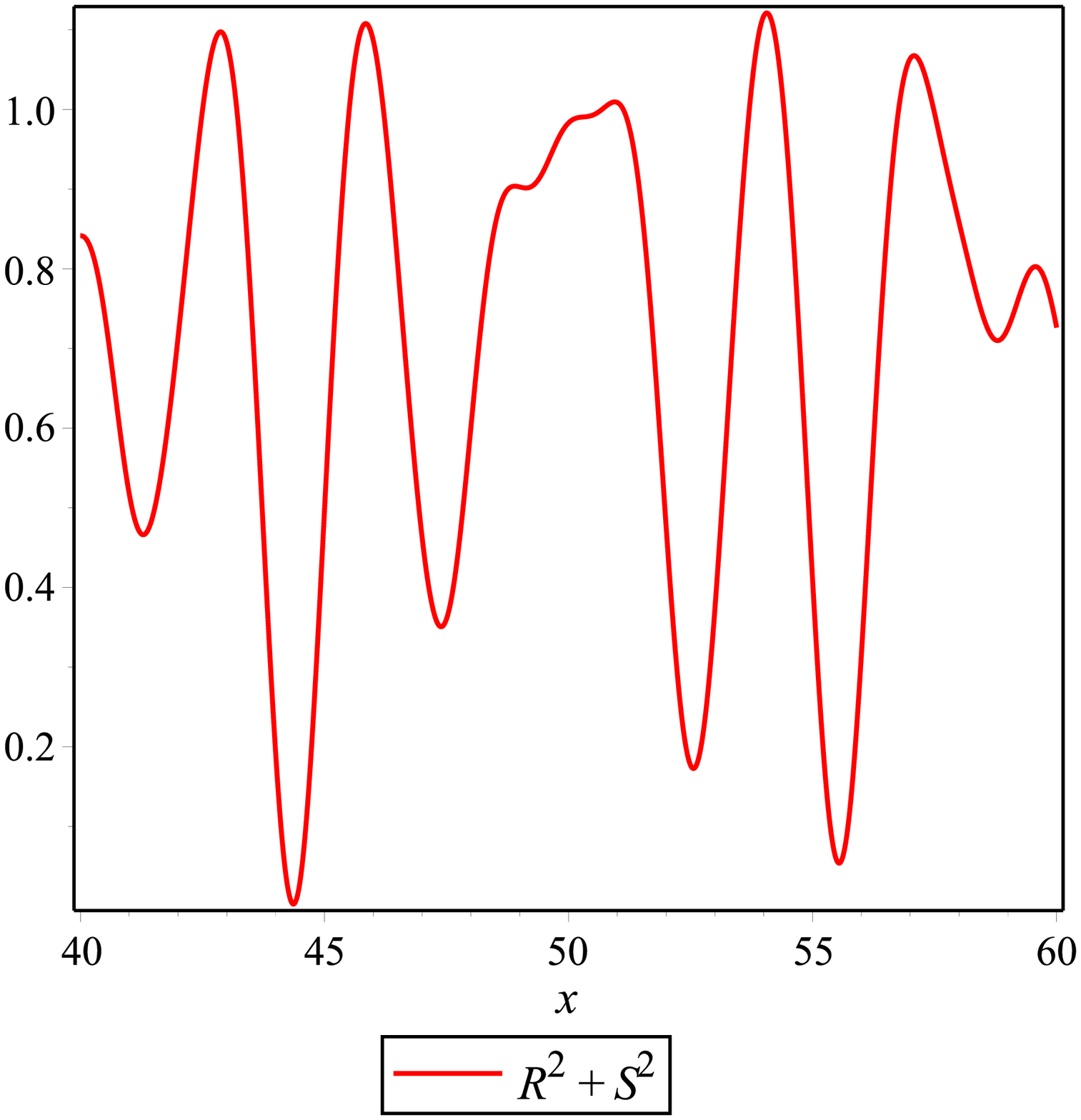}
$$
\text{(e)} \qquad\qquad\qquad\qquad\qquad\qquad \text{(f)}
$$
\vspace{-0.2in}
\includegraphics[width=0.3\textwidth]{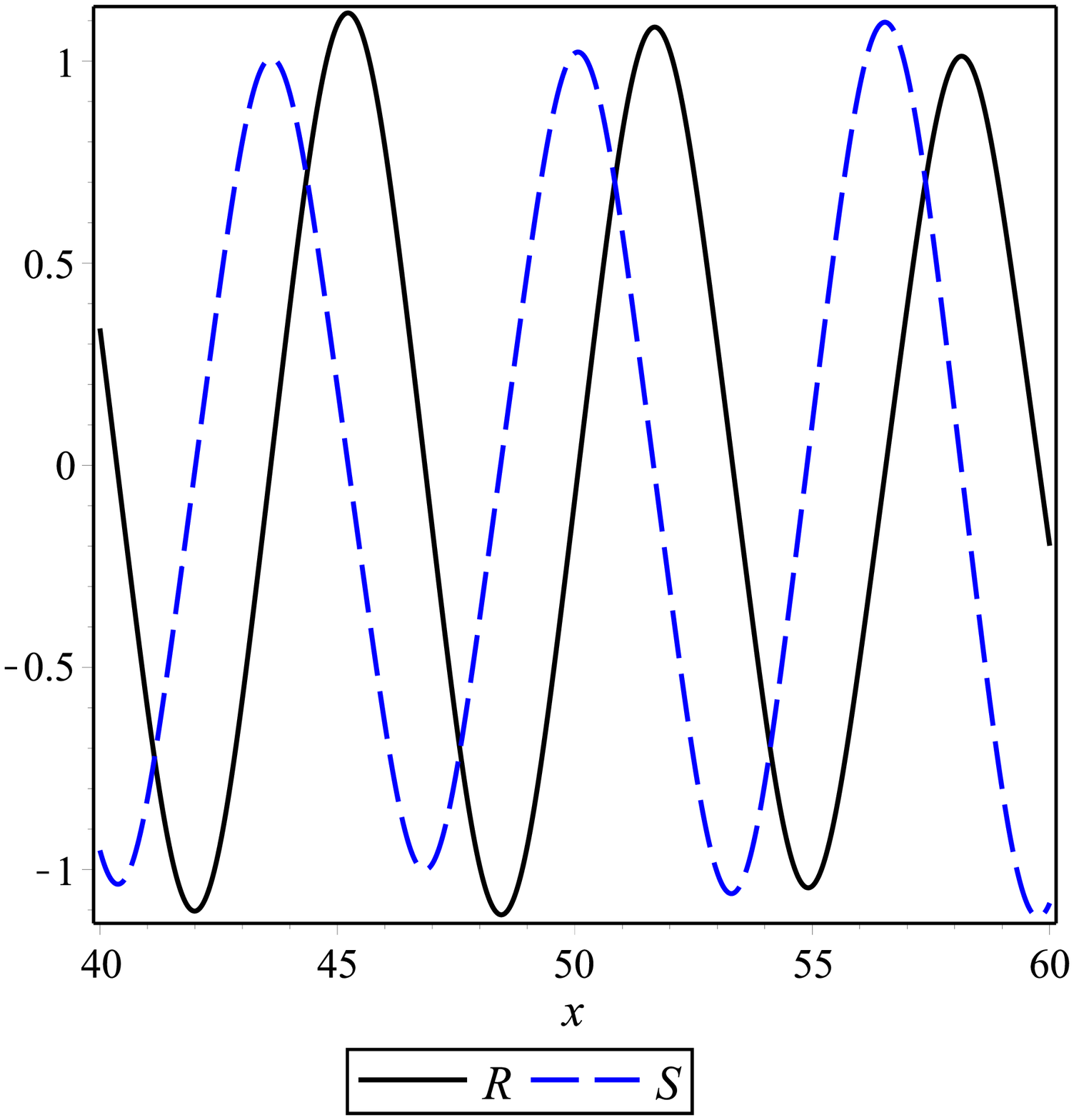}
\includegraphics[width=0.3\textwidth]{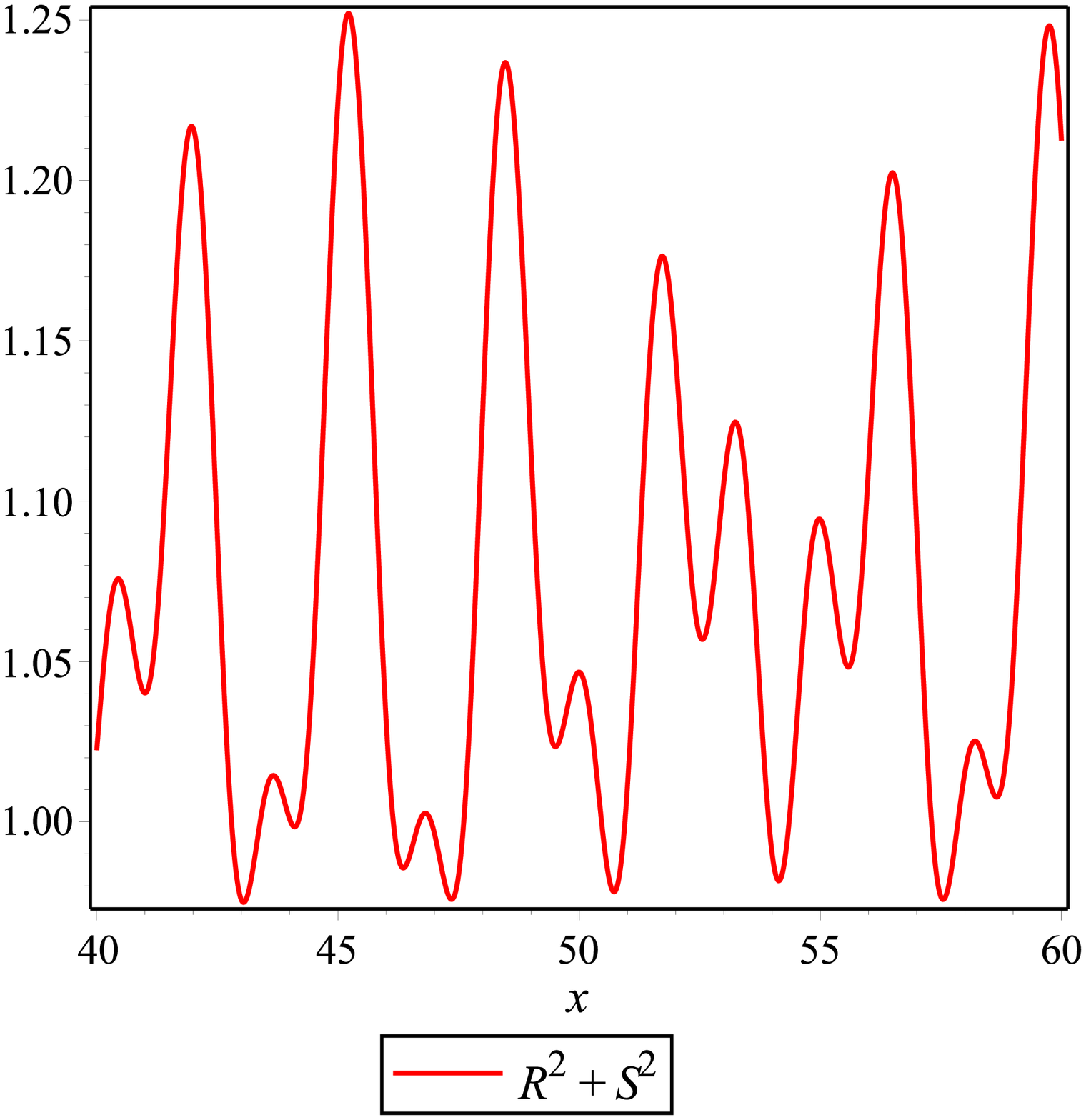}
$$
\text{(g)} \qquad\qquad\qquad\qquad\qquad\qquad \text{(h)}
$$
\vspace{-0.2in}
\caption{Plot of 1D stationary solutions $R$ and $S$ (left) and solution intensity $R^2 + S^2$ (right) governed by equations \eqref{ab3a}-\eqref{ab3b} for (a,b) $\alpha_1 = 0.1$, $\alpha_2 =1$, (c,d) $\alpha_1 = \alpha_2 =1$, (e,f) $\alpha_1 = 3$, $\alpha_2 =1$, (g,h) $\alpha_1 = \alpha_2 =2$. We set $\epsilon =2$, $\mu = 0.5$ and $n=2$, as well as $R(0)=1$, $R'(0)=0$, $S(0)=0$, $S'(0) =-1$ to obtain solutions which are out of spatial phase.}\label{figA}
\end{figure}

Through the above analysis, we have seen that aside from specific special cases, there do not appear to exist generic exact solutions to \eqref{ab3a}-\eqref{ab3b} when $\alpha_1 \neq \alpha_2$. Still, we can obtain numerical solutions, in order to obtain a better understanding of the solution behavior. These solutions are obtained via RKF45 implemented as part of the dsolve package in Maple 17. To this end, in Fig. \ref{figB} we plot stationary solutions for $n=1$ for various combinations of $\alpha_1$ and $\alpha_2$, given fixed conditions at $x=0$. We provide plots of the stationary solutions for the same choices of $\alpha_1$ and $\alpha_2$, only with $n=2$, in Fig. \ref{figA}.

In Figs. \ref{figB} and \ref{figA}, we find that for $\alpha_1 \neq \alpha_2$ there exist oscillatory yet non-periodic solutions. The intensity $R^2 + S^2$ of such solutions exhibits aperiodic dynamics, a further indication that such solutions are aperiodic. In the case where $\alpha_1 = \alpha_2 =1$, we obtain periodic solutions, however the intensity is non-constant. (We remark that, given different boundary conditions at $x=0$, one may tune the boundary conditions to obtain the exact solutions of \cite{chow2006exact}. Since this is not true of the asymmetric cases, we do not explore this further here.) The intensity is, however, periodic, and this suggests that even out-of-phase solutions exhibit a high degree of symmetry when both cross and self interaction terms are equal. When $\alpha_1 = \alpha_2$ yet both are not equal to unity, the solutions appear somewhat more ordered than their $\alpha_1 = \alpha_2$ counterparts, yet the intensity plots suggest that such solutions still lack periodicity. 

Our numerical solutions suggest that solutions corresponding to $\alpha_1 = \alpha_2 =1$ have the greatest degree of spatial symmetry (as suggested in previous sections, due to conditional integrability in this case), then these solutions lose symmetry for $\alpha_1 = \alpha_2 \neq 1$. Finally, when $\alpha_1 \neq \alpha_2$, the solutions exhibit more irregularity. This is again consistent with the aforementioned generic loss of integrability of the complex GL system for $\alpha_1 \neq \alpha_2$. 

Note that when the condition \eqref{ineq1} is broken, we can obtain (for certain parameter regimes considered) numerical solutions which do not remain bounded in space. 

\section{Numerical simulations and emergent dynamics}\label{numeric}
We numerically solve equations \eqref{sgl1}-\eqref{sgl2} using finite-differences in \textsc{Matlab}. We write four separate PDE for the real and imaginary parts of each wavefunction, and use centered differences to discretize the Laplacians, and solve the resulting large system of ODEs using the Runge-Kutta solver \textsc{`ODE45'}. We use absolute and relative tolerances of $10^{-9}$ and constrain the time step to confirm convergence of solutions. We pick suitably large domains of the form $[0,L]$ and $[0,L]^2$ with periodic boundary conditions to approximate $\mathbb{R}$ and $\mathbb{R}^2$ respectively. We take $L=10^3$ in all simulations. Unless stated otherwise, we always use random initial conditions in the interval $(0,0.1)$. In the 1-D setting we used $m=10^3$ grid points, and in the 2-D setting we used $m=400\times 400$ grid points.

\begin{figure}
\centering
\includegraphics[width=0.4\textwidth]{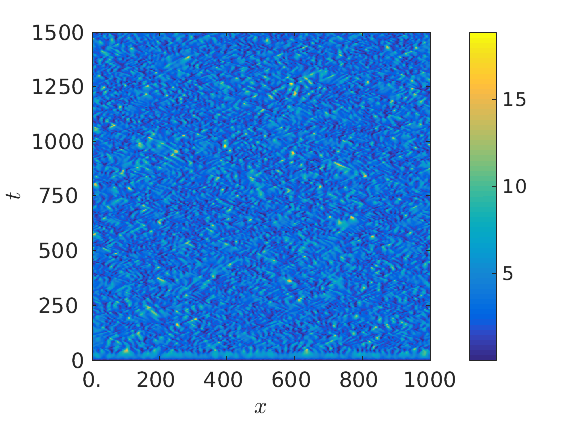}
\includegraphics[width=0.4\textwidth]{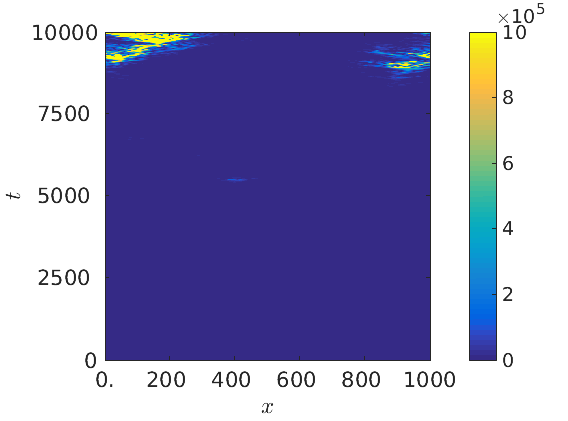}
\vspace{-0.2in}
$$
\text{(a)} \qquad\qquad\qquad\qquad\qquad\qquad\qquad\qquad\qquad \text{(b)}
$$
\vspace{-0.4in}
\caption{We plot the total intensity, $I = |u|^2+|v|^2$, for two values of $\epsilon$. In all plots we have taken $\alpha_1=\alpha_2=0.5$, $a=5$, $b=1$, $L=1000$, $n=2$, $\mu=0.7$, and then $\epsilon=0.896$ in (a), and $\epsilon=0.92$ in (b). Note that the solution is only bounded for $\epsilon < \bar{\epsilon} = 0.8964$. We take random initial data in $(0,1)$ for both plots. Panel (a) demonstrates a solution which remains bounded for all time, while in panel (b) we find that the solution grows without bound for large enough time, owing to the fact that the $\epsilon$ value corresponding to this solution is larger than $\bar{\epsilon}$.}\label{fig1}
\end{figure}

\subsection{Blow-up and boundedness}
We numerically confirm the boundedness of solutions consistent with the results approximating the absorbing set in Sec. \ref{sec3.1}. For all values of $\alpha_1+\alpha_2 \geq 2$ we find that solutions remain bounded for large time for all parameter ranges we explored if $\epsilon < \bar{\epsilon}$. This bound is sharp, as for $\epsilon > \bar{\epsilon}$, random small initial data grow arbitrarily large and hence appear to become unbounded. This is a very rapid and spatially uniform process, in that within $1000$ units of time $t$ the intensity is larger than $10^{300}$. In the other case, where $\alpha_1+\alpha_2 < 2$, we again confirm boundedness for $\epsilon < \bar{\epsilon}$, and unbounded solutions for $\epsilon \gg \bar{\epsilon}$, but there are more complicated behaviours for $\epsilon \gtrapprox \bar{\epsilon}$. We show an example of this in Fig. \ref{fig1}, where bounded initial data remain so for $\epsilon < \bar{\epsilon}$. When this bound is exceeded, however, small pockets of large intensity appear as time progresses, and these appear to become larger over time, growing without bound. We confirm this by increasing the magnitude of the initial data, our simulations showing an increase in the magnitude of the spikes later in time.

\subsection{Separation of intensities}
\begin{figure}
\centering
\includegraphics[width=0.4\textwidth]{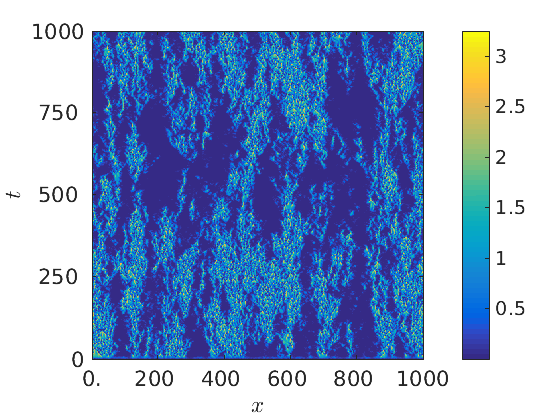}
\includegraphics[width=0.4\textwidth]{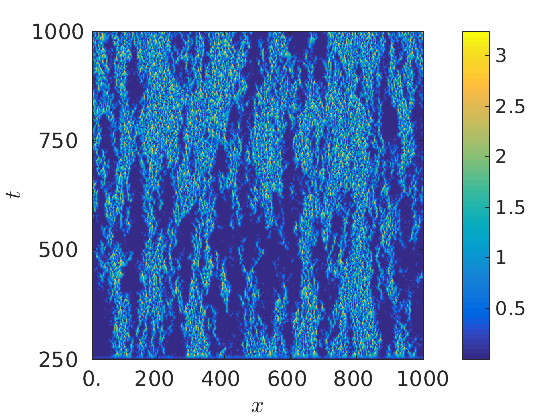}
\vspace{-0.2in}
$$
\text{(a)} \qquad\qquad\qquad\qquad\qquad\qquad\qquad\qquad\qquad \text{(b)}
$$
\includegraphics[width=0.4\textwidth]{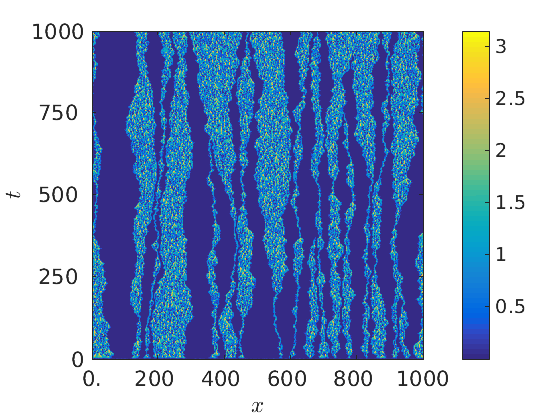}
\includegraphics[width=0.4\textwidth]{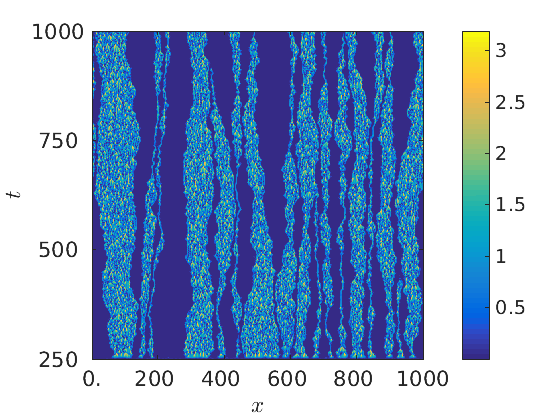}
\vspace{-0.2in}
$$
\text{(c)} \qquad\qquad\qquad\qquad\qquad\qquad\qquad\qquad\qquad \text{(d)}
$$
\vspace{-0.4in}
\caption{Plots of $I_u$ on the left and $I_v$ on the right for increasing $\alpha_1=\alpha_2$. In all plots we have taken $\epsilon=1$, $a=2$, $b=2$, $L=1000$, $n=2$, $\mu=0.7$, and then $\alpha_1=\alpha_2=1.5$ in (a)-(b) and $\alpha_1=\alpha_2=5$ in (c)-(d). In each case, we observe a separation of intensity of the two wave functions. As the XPM parameters$\alpha_1$ and $\alpha_2$ increase in value, we find that the solutions tend to segregate within bands. Within a band, the dynamics appear to give spatiotemporal chaos, which verifies the fact that modulational instability of wavefunctions appears generic to the system.}\label{fig3}
\end{figure}

In this section, we demonstrate an interesting effect for large XPM (that is, when $\alpha_i > 1$) whereby the intensity of each wavefunction, defined as $I_u = |u|^2$ and $I_v = |v|^2$ respectively, become spatially separated. That is, for large times and for $\alpha_i \gtrapprox 1.4$, we always observe that each respective intensity is nonzero only when the other is approximately zero. In Fig. \ref{fig3} we plot these corresponding intensities for three levels of $\alpha_1=\alpha_2$ in each plot to demonstrate this effect. For $\alpha_1 = \alpha_2 =1$, plots instead appear like that in Fig. \ref{fig1}(a).

\begin{figure}
\centering
\includegraphics[width=0.4\textwidth]{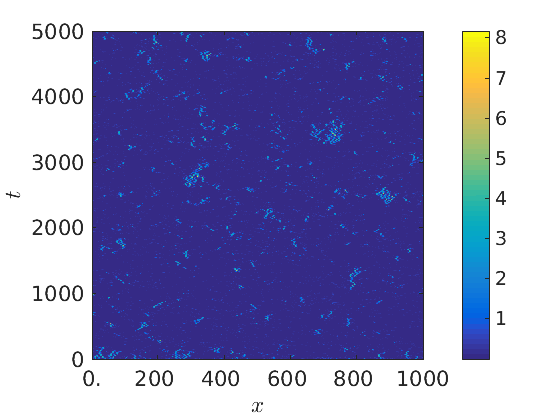}
\includegraphics[width=0.4\textwidth]{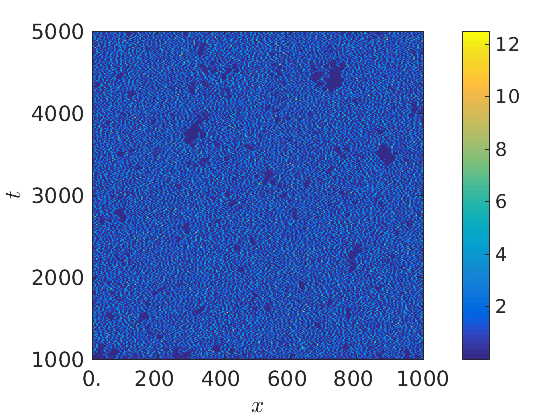}
\vspace{-0.2in}
$$
\text{(a)} \qquad\qquad\qquad\qquad\qquad\qquad\qquad\qquad\qquad \text{(b)}
$$
\includegraphics[width=0.4\textwidth]{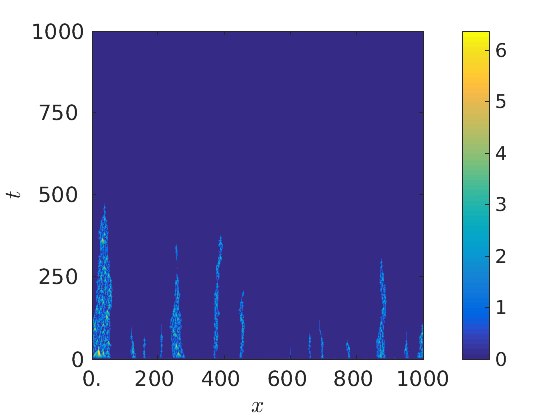}
\includegraphics[width=0.4\textwidth]{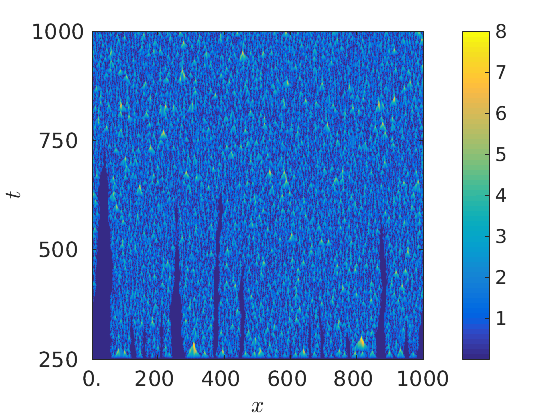}
\vspace{-0.2in}
$$
\text{(c)} \qquad\qquad\qquad\qquad\qquad\qquad\qquad\qquad\qquad \text{(d)}
$$
\vspace{-0.4in}
\caption{Plots of $I_u$ on the left and $I_v$ on the right for different $\alpha_1 \neq \alpha_2$. In all plots we have taken $\epsilon=1$, $L=1000$, $n=2$, and then fix $\alpha_1=1.5$, $\alpha_2=1.4$, $a=5$, $b=5$ and $\mu=0.99$ in (a)-(b), $\alpha_1=5$, $\alpha_2=4$, $a=1$, $b=3$ and $\mu=0.99$ in (c)-(d). When XPM parameters are different but still close in value, one wavefunction is preferred, while one is confined to a more restricted region. However, as panel (a) demonstrates, this restricted region can itself appear rather disordered. For greater differences between XPM, conditions \eqref{die1}-\eqref{die2} may come into play, resulting in extinction of one of the wavefunctions, as seen in panel (c) by time $t=500$. These simulations therefore give further support for the analytical results in Sec. \ref{secloal}.} \label{fig4}
\end{figure}

Within these regions of nonzero intensity (or in the case of equal intensities where each wavefunction is mixed) the generic behave of the intensity is a kind of spatiotemporal chaos. In some cases, this induces a meandering boundary of the support of the wavefunction. We note that some regions that have nonzero intensity for one wavefunction can disappear due to this chaotic motion changing the boundary of this region; see for instance, Fig. \ref{fig4}.

If $\alpha_1 \neq \alpha_2$, and one of the conditions \eqref{die1} or \eqref{die2} holds, then the equation containing the larger XPM parameter will be driven to zero. We give examples of this in Fig. \ref{fig4}. In the first example (corresponding to $\alpha_1=1.5$, $\alpha_2=1.4$), the XPM parameters are not large enough to fully separate the wavefunctions, and hence $I_u$ persists in sporadic pockets over time. In the second example (corresponding to $\alpha_1=5$, $\alpha_2=4$), $I_u$ is driven to zero as $I_v$ tends to a spatiotemporal chaos state throughout the domain. We observe that some `tendrils' of $u$ extending in time, corresponding to small regions of nonzero $I_u$ that persist in time, but these are eventually extinguished by fluctuations of the larger wavefunction.

\subsection{Comparison to the cubic complex Ginzburg-Laundau system}
We now vary $\mu$ to compare the behavior of the system with saturable kinetics to that of the cubic complex GL system. We note that the boundedness result for $\epsilon < 1/\mu^{1/n}$ (assuming $\alpha_1+\alpha_2>2$), implies that for a given $\epsilon$, solutions will be bounded for $\mu < 1/\epsilon^n$, and we exploit this to consider large values of $\mu$ in the saturable model. 

\begin{figure}
\centering
\includegraphics[width=0.4\textwidth]{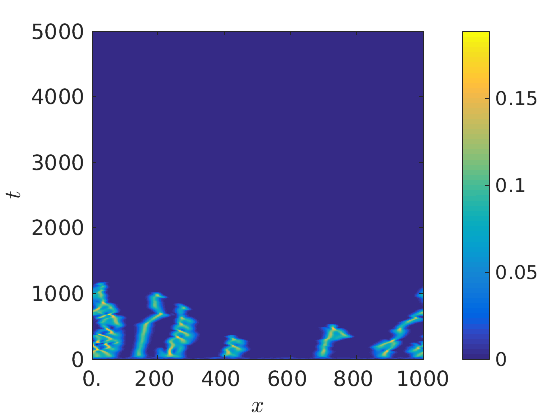}
\includegraphics[width=0.4\textwidth]{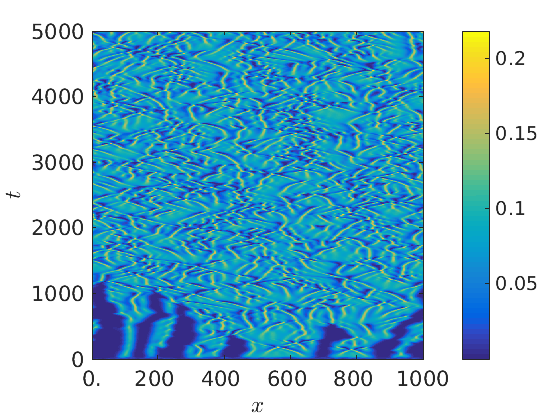}
\vspace{-0.2in}
$$
\text{(a)} \qquad\qquad\qquad\qquad\qquad\qquad\qquad\qquad\qquad \text{(b)}
$$
\includegraphics[width=0.4\textwidth]{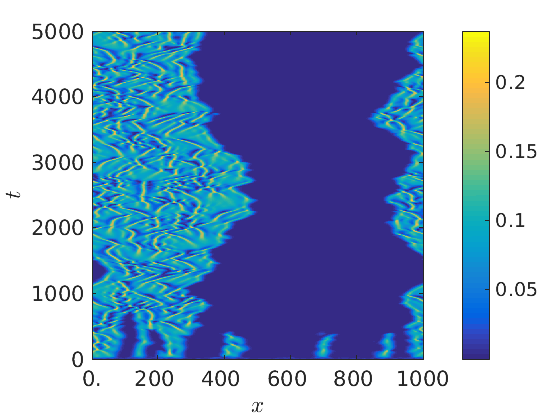}
\includegraphics[width=0.4\textwidth]{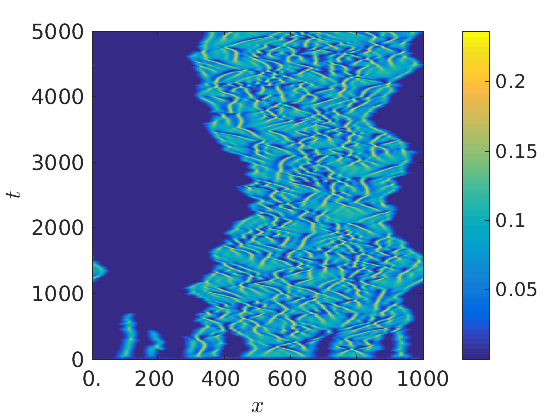}
\vspace{-0.2in}
$$
\text{(c)} \qquad\qquad\qquad\qquad\qquad\qquad\qquad\qquad\qquad \text{(d)}
$$
\includegraphics[width=0.4\textwidth]{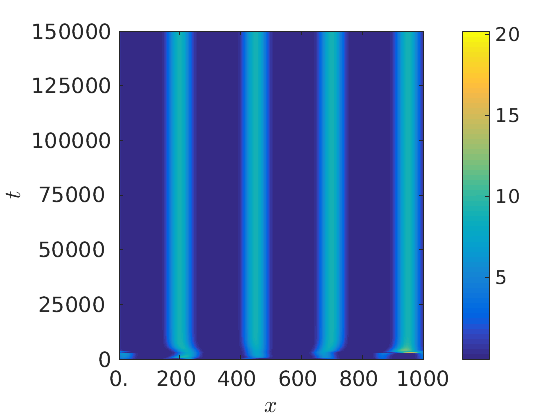}
\includegraphics[width=0.4\textwidth]{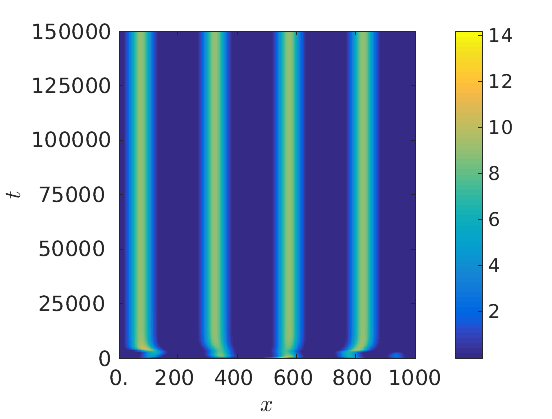}
\vspace{-0.2in}
$$
\text{(e)} \qquad\qquad\qquad\qquad\qquad\qquad\qquad\qquad\qquad \text{(f)}
$$
\vspace{-0.4in}
\caption{Plots of $I_u$ on the left and $I_v$ on the right for increasing $\mu$. In all plots we have taken $\epsilon=0.099$, $L=1000$, $n=1$, and then $\alpha_1=5$, $\alpha_2=5$, $a=5$, $b=1$ and $\mu=0$ in (a)-(b), $\mu=1$ in (c)-(d), $\mu=10$ in (e)-(f). Note the variation in timescales. For $\mu =0$, we recover the cubic complex GL system, and with the choice of different XPM parameters, we have the extinction of the $u$ wavefunction for large enough time. As $\mu$ increases, the solutions are in a sense regularized, and the two wavefunctions are seen to both persist for $\mu =1$. The dynamics of both $\mu =0$ and $\mu=1$ appear to exhibit spatiotemporal chaos. Interestingly, when $\mu$ becomes quite large (here we set $\mu =10$), the dynamics segregate into discrete bands for large time, with these structures appearing to become stable. The amplitude of these solutions is large (relative to the other cases) indicating that the solutions grow in amplitude and become self-reinforcing against perturbations when saturability of the media is strong, for our choice of $a$ and $b$.}\label{fig5}
\end{figure}

We use increasing values of $\mu$ and plot intensities $I_u$ and $I_v$ in Fig. \ref{fig5}. Generally, the differences between the saturable dynamics and the $\mu=0$ cubic case depends on how close, in a relative sense, the parameters are to the blow-up boundary of $\bar{\epsilon}$. We note that the timescales are taken larger with $\mu$, in order to observe the behavior. Similarly, the maximum intensity of each plot increases as $\mu$ increases. We note qualitatively different transient behaviors occur for changes in the saturation parameter $\mu$. 

\begin{figure}
\centering
\includegraphics[width=0.4\textwidth]{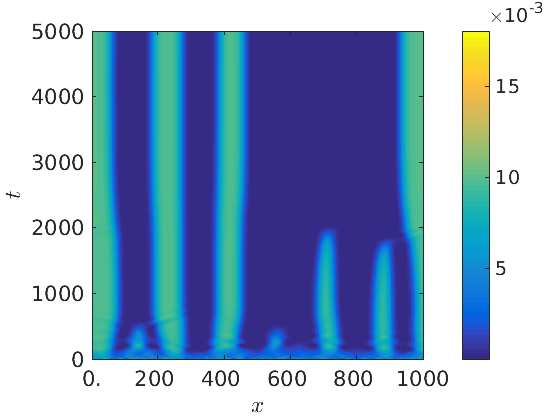}
\includegraphics[width=0.4\textwidth]{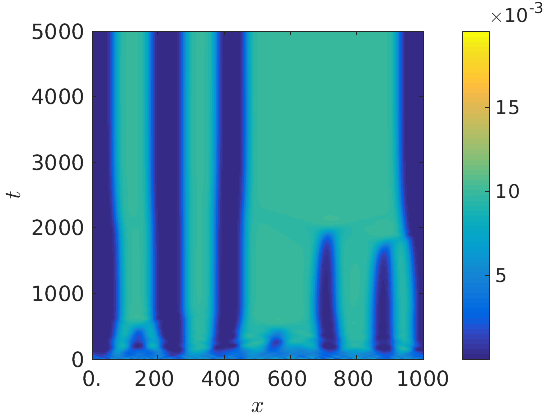}
\vspace{-0.2in}
$$
\text{(a)} \qquad\qquad\qquad\qquad\qquad\qquad\qquad\qquad\qquad \text{(b)}
$$
\includegraphics[width=0.4\textwidth]{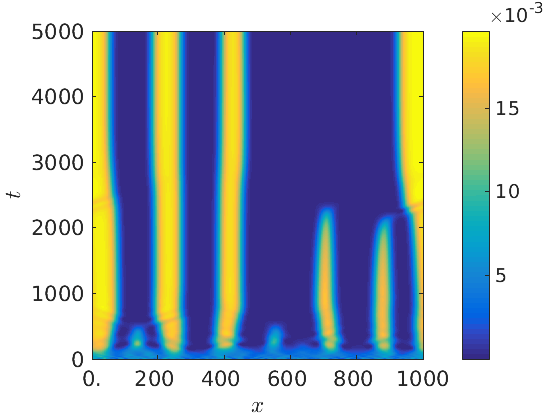}
\includegraphics[width=0.4\textwidth]{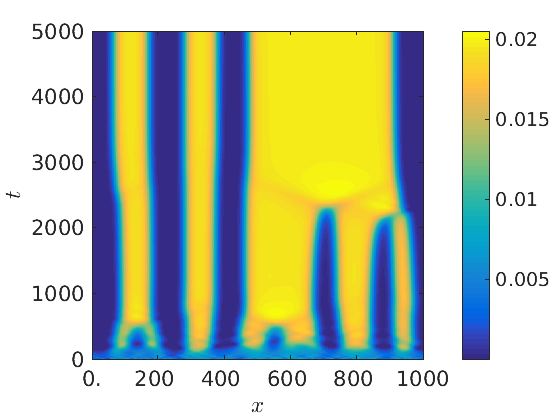}
\vspace{-0.2in}
$$
\text{(c)} \qquad\qquad\qquad\qquad\qquad\qquad\qquad\qquad\qquad \text{(d)}
$$
\includegraphics[width=0.4\textwidth]{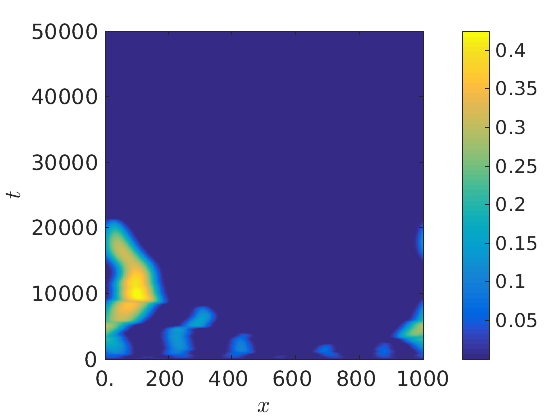}
\includegraphics[width=0.4\textwidth]{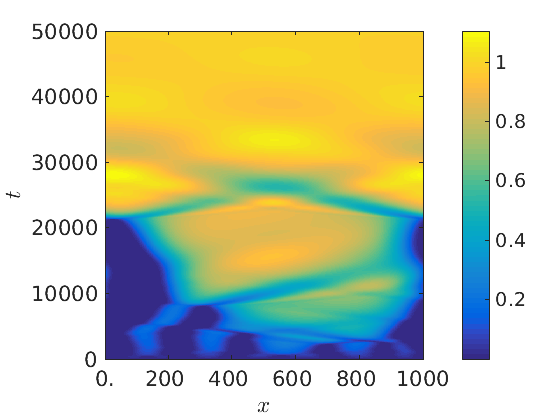}
\vspace{-0.2in}
$$
\text{(e)} \qquad\qquad\qquad\qquad\qquad\qquad\qquad\qquad\qquad \text{(f)}
$$
\vspace{-0.4in}
\caption{Plots of $I_u$ on the left and $I_v$ on the right for increasing $\mu$. In all plots we have taken $\epsilon=0.0099$, $L=1000$, $n=1$, and then $\alpha_1=2$, $\alpha_2=2$, $a=-5$ (note the sign), $b=5$, while $\mu=0$ in (a)-(b), $\mu=50$ in (c)-(d), and $\mu=100$ in (e)-(f). For these parameters (in particular, for $a<0$), we find that small values of $\mu$ are somewhat stabilizing, resulting in banding patters akin to those found in Fig. \ref{fig5}. However, these bands are fairly irregular in appearance, with some bands eventually merging at large time. This is in contrast to the discrete nature of the bands observed in (e)-(f) of Fig. \ref{fig5}. Increasing $\mu$ from $\mu =0$ to $\mu =50$, we find that the overall structure of the solutions is the same, with the only qualitative change being an increase in the amplitude of the solutions. However, when we increase $\mu$ further, there is a bifurcation into a new regime, with initial attempts at banding instead dying out (e) or merging into a single region (f). This provides an example of a parameter regime where increasing the saturability in the model actually results in a loss of regularity in the discrete banded structures, with one of the wavefunctions dissipating completely for large time.}\label{fig6}
\end{figure}

We note that for large values of the XPM, for some parameter regimes, we observe one of the wavefunctions becoming extinct (as in the case with unequal $\alpha_i$). In Fig. \ref{fig5}, the top left panel exhibits such behavior, as does a panel in Fig. \ref{fig6}. We note that this is not unique to saturable kinetics; the top left panel of Fig. \ref{fig5} corresponds to the purely cubic kinetics. However, since the complex GL system with cubic nonlinearity is not usually studied with general XPM parameters, we are not aware of this being pointed out previously in the literature.

We observe that larger values of $\mu$ tend to smooth and coarsen the dynamics, or in some cases regularize them as in Fig. \ref{fig5} where for $\mu=10$, both intensities tend to steady but simply-patterned states, whereas for smaller values of $\mu$ the intensities behave as more generic spatiotemporal fluctuations. A complete classification of when the dynamics coalesce to form steady intensity profiles is not tractable, but we do note that this behavior is extremely rare if $a$ and $b$ have the same sign (with spatiotemporal chaos appearing far more often in simulations), but common if they have opposing signs as in Fig. \ref{fig6}. Note also that, for $a$ and $b$ of opposing signs, we observe that while there is still a coarsening of the dynamics for large saturability, there may be loss or extinction of one of the wavefunctions, as seen in Fig. \ref{fig6}(e). Therefore, the coexistence of smooth, temporally steady patterns depends strongly on not only the saturability of the media, but on the parameters $a$ and $b$, as well. 
 
We comment that the initially periodic structures observed in Fig. \ref{fig6} are consistent with the stationary solutions discussed in Sec. \ref{sec2e}. In particular, for Fig. \ref{fig6} we have $a+b=0$. In some cases, the solutions transiently take on the form of a spatially periodic profile (or, more generally, a spatially oscillatory profile), before losing stability and breaking up into other structures. In some cases, transient periodic structures break into spatially oscillatory yet non-periodic structures for large time. This is akin to what was observed in Figs. \ref{figB} and \ref{figA}, where non-periodic oscillatory structures were the generic form of a bounded stationary solution. Of course, we did not study the stability of stationary solutions as a function of the model parameters, however what we are likely observing in our simulations is the stability of certain heterogeneous stationary solutions, and the instability of others.

\subsection{Localized structures}
\begin{figure}
\centering
\includegraphics[width=0.4\textwidth]{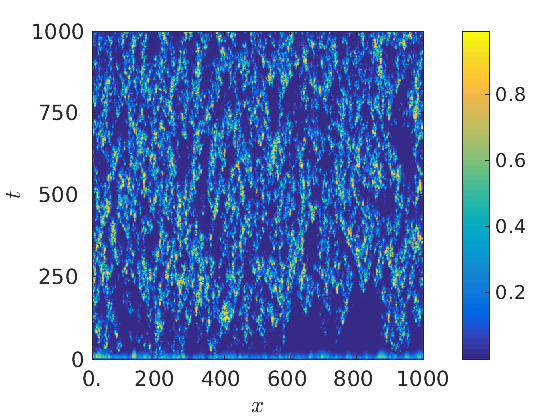}
\includegraphics[width=0.4\textwidth]{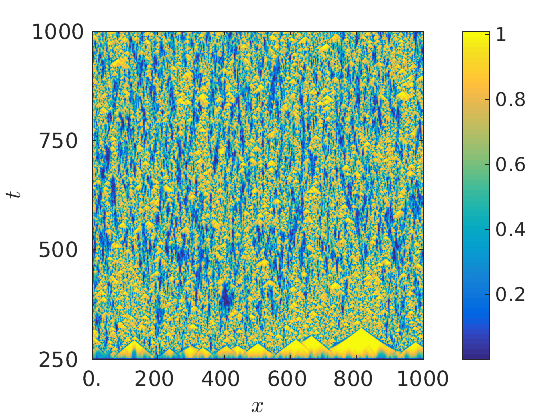}
\vspace{-0.2in}
$$
\text{(a)} \qquad\qquad\qquad\qquad\qquad\qquad\qquad\qquad\qquad \text{(b)}
$$
\includegraphics[width=0.4\textwidth]{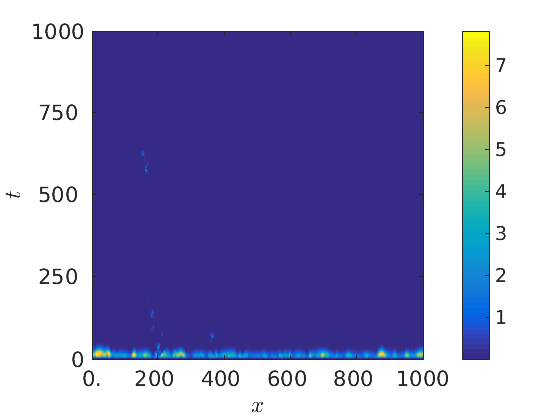}
\includegraphics[width=0.4\textwidth]{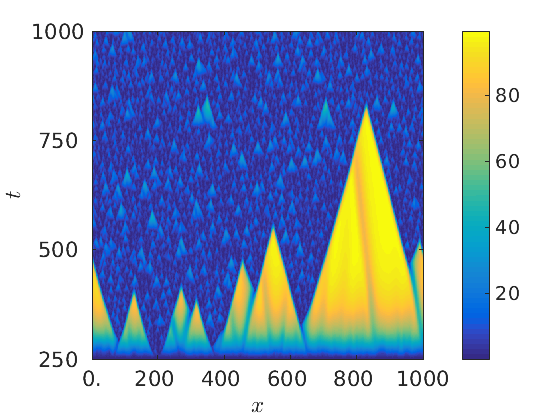}
\vspace{-0.2in}
$$
\text{(c)} \qquad\qquad\qquad\qquad\qquad\qquad\qquad\qquad\qquad \text{(d)}
$$
\vspace{-0.4in}
\caption{Plots of $I_u$ on the left and $I_v$ on the right for increasing $\mu$. In all plots we have taken $\epsilon=1$, $L=1000$, $n=1$, and then $\alpha_1=1$, $\alpha_2=1.5$, $a=0.1$, $b=5$ and $\mu=0.01$ (a)-(b), $\mu=0.99$ (c)-(d). In each case we observe transient defects of triangular form, although these are more pronounced in cases of extinction of one of the wavefunctions, as seen in (c)-(d) where $\mu$ is taken to be larger. }\label{fig9}
\end{figure}

In addition to spatiotemporal chaos or regular steady state patterns such as bands or even uniform states, there are other emergent dynamics possible from \eqref{sgl1}-\eqref{sgl2} in restricted parameter regimes, at least at intermediate timescales. In Fig. \ref{fig9}, we observe transient defects in the intensity plots that eventually deteriorate into spatiotemporal chaos. However, as we increase $\mu$, we see these transient defects persist for longer periods of time, again emphasizing the kind-of regularizing nature of being closer to the blow-up parameter boundary. The triangular shaped transient defects have been observed previously in the literature on vector cubic GL systems \cite{hernandez1999spatiotemporal}.

Note that the transient defects are more prevalent when $\mu$ is large. So, not only do the defects persist for longer, they also are a more dominant proportion of the dynamics. For small $\mu$ the transient defects will give way to spatiotemporal chaos, while when $\mu$ is larger they may tend to uniform or existence states. For larger $\mu$ we also observe a resurgence of new transient defects (see Fig. \ref{fig9}(d)), with new defects forming after the initial defects have dissipated. This is in contrast to the small-$\mu$ limit (which includes the cubic complex GL system), where the initial transient defects will give way to spatiotemporal chaos. This again suggests one role of $\mu$ is to regularize the dynamics of the saturable GL system.

\begin{figure}
\centering
\includegraphics[width=0.4\textwidth]{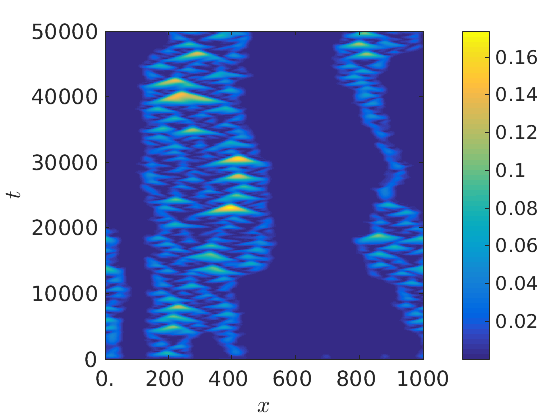}
\includegraphics[width=0.4\textwidth]{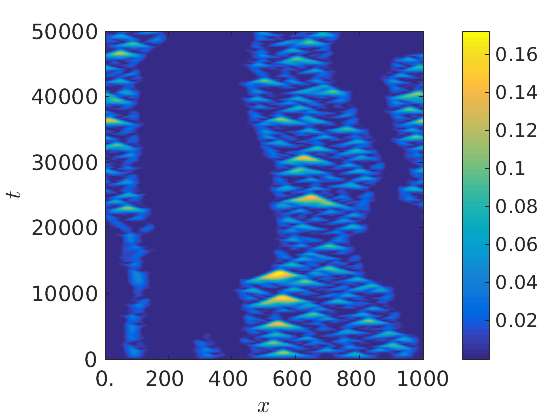}
\vspace{-0.2in}
\caption{Plots of $I_u$ on the left and $I_v$ on the right for $\epsilon=0.0099$, $L=1000$, $\alpha_1=5$, $\alpha_2=5$, $a=0$, $b=5$, $n=1$, and $\mu=100$. Within each band there appear to be highly localized waves with peak amplitude much larger than that of surrounding waves. The localization is in space and in time, with the waves pulsing several times before dissipating. As the amplitude of these waves is up to and greater than three times the mean amplitude of surrounding waves, these localized structures may be candidates for transient rogue waves.  }\label{fig7}
\end{figure}

Finally, we note the existence of what appear to be rogue waves - isolated spatial maxima of amplitude greater than three times the surrounding wave height - that occur within pockets of spatiotemporal chaos. We note examples of these in Fig. \ref{fig7}, where these highly localized waves occur as bursts in similar spatial regions to where they occurred previously. While they share some features of breathers, they exhibit far less regularity and appear to always die out after sufficient time. It is interesting to see such dynamics even when saturation is very large, as an increase in $\mu$ often results in a stabilization of the dynamics. Optical rogue waves were previously reported in higher-order (fourth-order, in particular) complex scalar GL equations \cite{gibson2016optical}, and so the emergence of rogue waves in certain parameter regimes for our model may not be surprising.

\subsection{Dynamics and pattern formation in two spatial dimensions}
\begin{figure}
\centering
\includegraphics[width=0.4\textwidth]{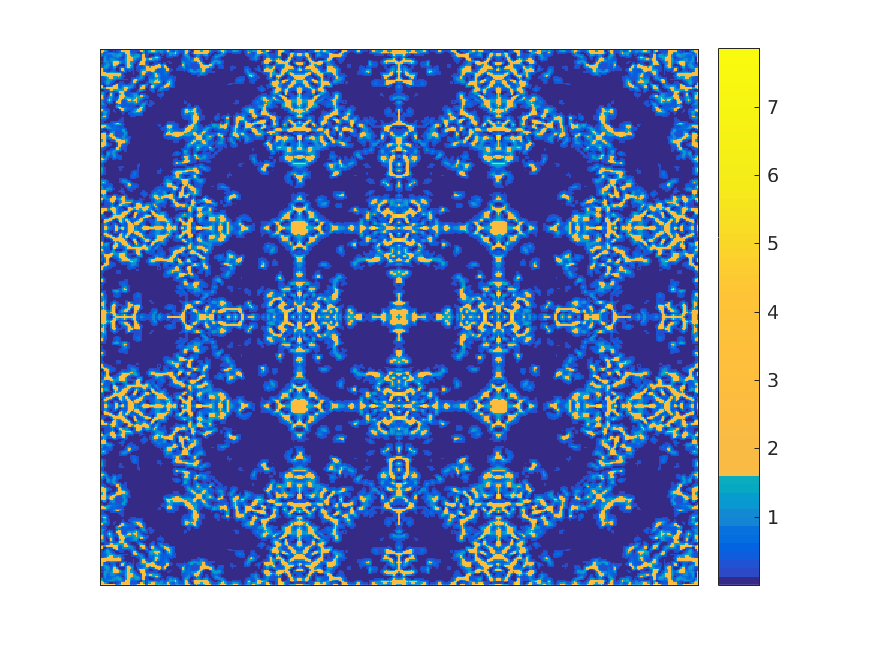}
\includegraphics[width=0.4\textwidth]{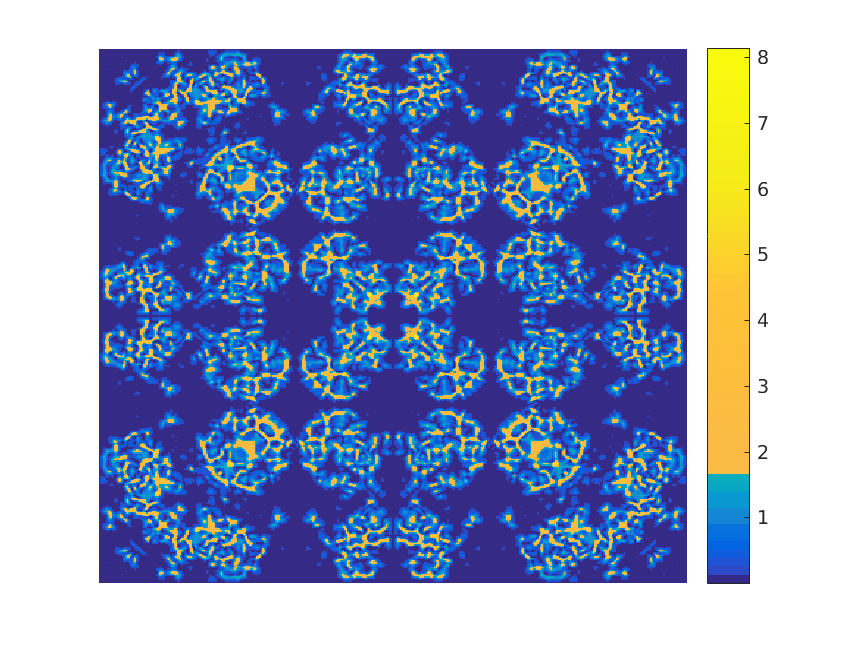}
\includegraphics[width=0.4\textwidth]{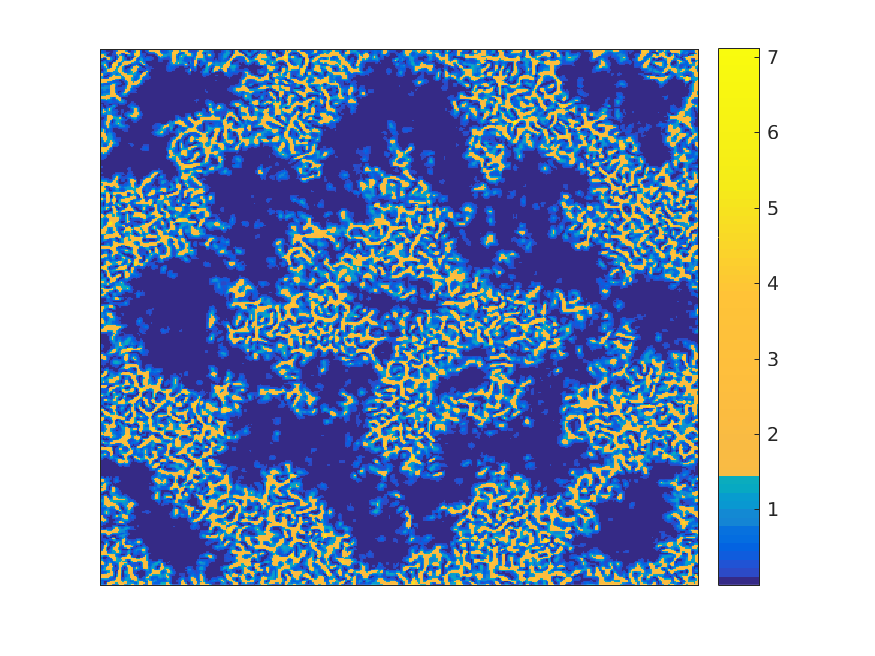}
\includegraphics[width=0.4\textwidth]{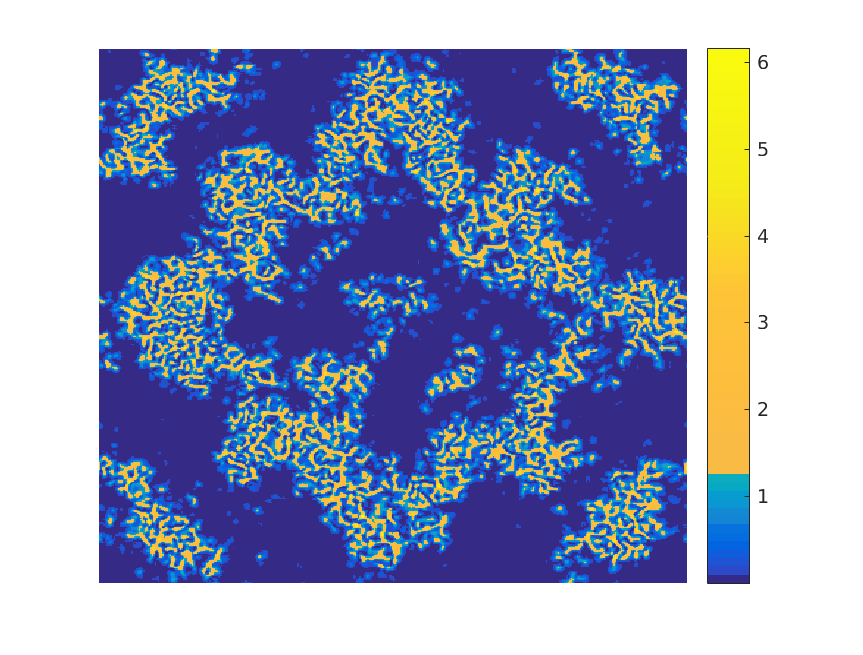}
\vspace{-0.15in}
\caption{Plots of $I_u$ on the left and $I_v$ on the right for $t=130$ (top) and $t=250$ (bottom). In all plots we have taken $\alpha_1=\alpha_2=2$, $a=5$, $b=1$, $L=1000$, $n=2$, $\mu=0.7$, and $\epsilon=1$. }\label{fig10}
\end{figure}

\begin{figure}
\centering
\includegraphics[width=0.4\textwidth]{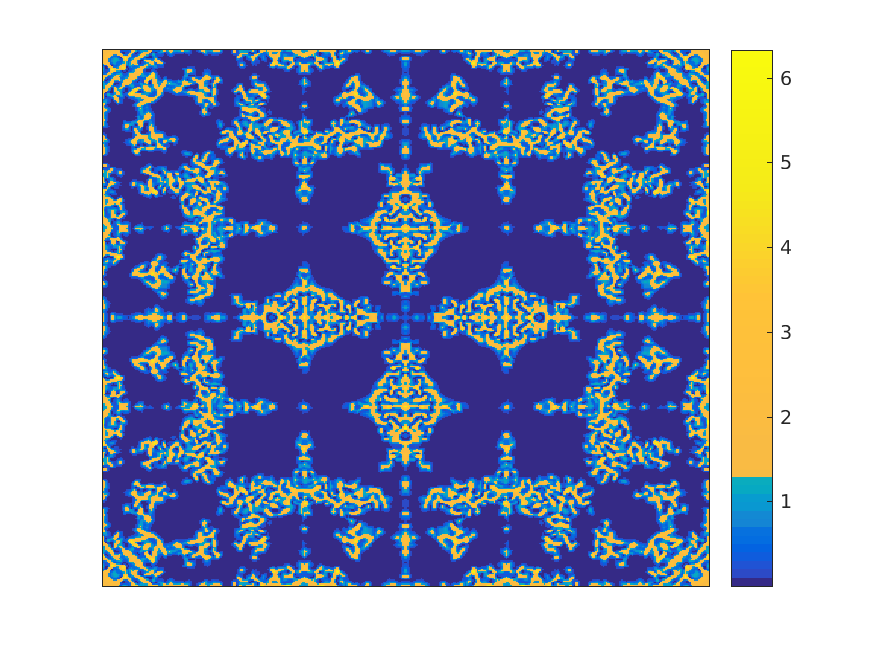}
\includegraphics[width=0.4\textwidth]{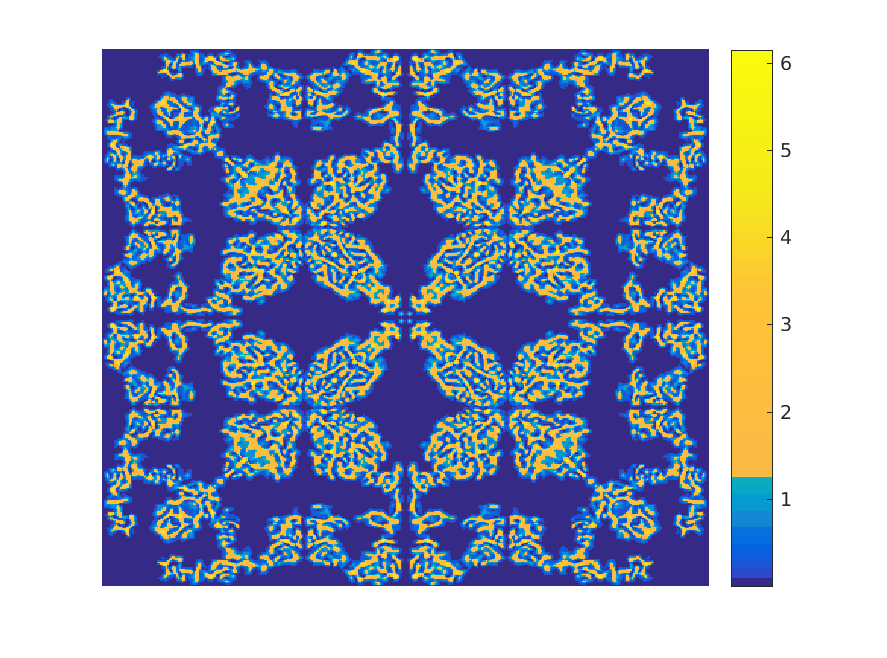}
\includegraphics[width=0.4\textwidth]{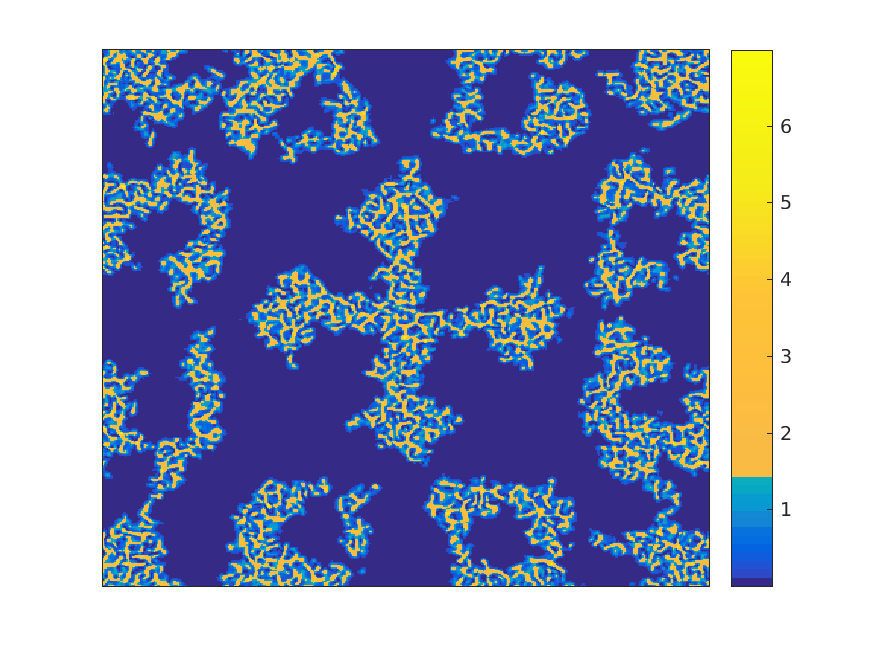}
\includegraphics[width=0.4\textwidth]{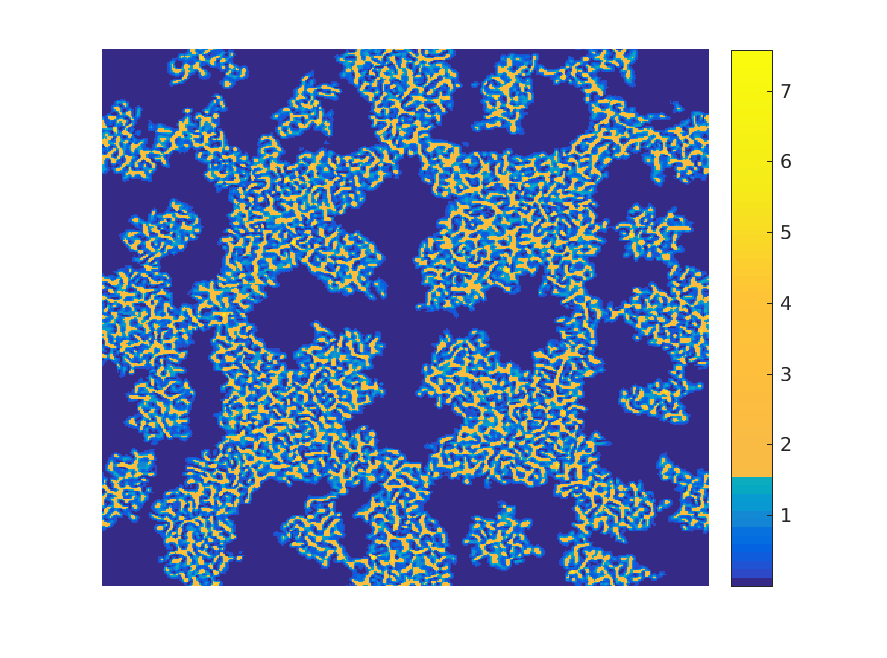}
\vspace{-0.15in}
\caption{Plots of $I_u$ on the left and $I_v$ on the right for $t=130$ (top) and $t=250$ (bottom). In all plots we have taken $\alpha_1=\alpha_2=5$, $a=5$, $b=1$, $L=1000$, $n=2$, $\mu=0.7$, and $\epsilon=1$.}\label{fig11}
\end{figure}

Finally we demonstrate patterns observed in two-dimensional spatial domains with specific initial data. Generic behavior for small random initial conditions corresponds to random fluctuations as we have seen in the 1-D case, and we also remark that for small $\mu$ we can recover many kinds of patterns found in the cubic complex GL system in the literature. Here we give an example of spontaneous patterning arising from an initial seed of $u(0,x,y) = (1+i)\cos(x)\cos(y)$ and $v(0,x,y) = (1+i)\sin(x)\sin(y)$. In Fig. \ref{fig10} we demonstrate the evolution of this initial data for $\alpha_1=\alpha_2=2$, and in Fig. \ref{fig11} for $\alpha_1=\alpha_2=5$. Note that the color bars have been scaled to emphasize regions of high-intensity.

As we might anticipate from the one-dimensional simulations, for $\alpha_1=\alpha_2=1$, patterns form and mingle between the two wavefunctions, without any clear separation between them (although we do notice that they pattern differently, likely corresponding to interactions between them). For larger XPM parameters, we see separation between the two wavefunction intensities as they form intricate patterns on the boundaries between high intensity areas. Over long periods of time, these patterns break down into spatiotemporal fluctuations as expected, but they seem to persist in a dynamic way (e.g. continually changing and shifting) for long periods of time depending on parameters chosen.

\section{Discussion}\label{dis}
We studied the dynamics emergent from a complex Ginzburg-Landau system with saturable nonlinearity including cross-phase modulation (XPM) parameters. First studying the system analytically, we obtained conditions for the existence of bounded dynamics. One interesting result was the amplitude death of individual wavefunctions in certain parameter limits, which gives a nice specific instance of general results in \cite{van2018generic}. In particular, we can give the amplitude death conditions in a way which is akin to local stability of a steady state, although one can check that our result is in agreement with those of \cite{van2018generic} (which make use of inverse functions of the monotone nonlinear terms) in a neighborhood of the amplitude death states. We were also able to construct general plane wave solutions in closed form, and using these we were able to demonstrate generic modulational instability of such plane wave solutions for many combinations of the model parameters given small enough wavenumber perturbations, suggesting that unsteady dynamics (such as spatiotemporal chaos) should prevail when bounded solutions exist. We also considered the degenerate limit $a+b=0$, which corresponds to the limit where the complex GL dynamics reduce to those of a coupled saturable NLS system. While the resulting reduced system is still not generically integrable (save for the case where cross and self interaction terms are equal), we were able to study a variety of spatially heterogeneous stationary solutions.

Many of the aforementioned dynamics were later demonstrated numerically. Regarding numerical simulations, for XPM parameters much larger than self-interaction parameters, we found from numerical simulations that segregation of wavefunctions into different regions of the spatial domain was a common occurrence. Within each respective region, the underlying dynamics for XPM parameters set to unity were found, with behaviors ranging from steady uniform dynamics or patterning to spatiotemporal chaos depending on the values of other model parameters. We found a variety of transient dynamics which persist for intermediate times, including transient defects (such as those of triangular form shown in Fig. \ref{fig9}) and spatial patterns in two-dimensional spatial domains (as shown in Figs. \ref{fig10}-\ref{fig11}). When XPM parameters were large and yet of distinct values, one wavefunction was seen in simulations to decay to zero in finite time over the spatial domain. In order to better understand these numerical results, we obtained analytic conditions for amplitude death of one of the wavefunctions in the presence of unequal XPM parameters, and we found amplitude death of $u$ for large enough $\alpha_1$ and amplitude death of $v$ for large enough $\alpha_2$.

The size and extent of the separate regions (found when XPM are taken to be larger than the self-interaction parameters) changes with time, and the support of one wavefunction may occupy a larger region of the spatial domain at one time, only for the support of the other wave function to dominate at a later time. Within the support of each wavefunction, spatiotemporal chaos is often found (see, for instance, Fig. \ref{fig3}), although for some specific parameter regimes localized structures or stable patterns may emerge; see Fig. \ref{fig5}. For large enough XPM parameters, these bands will occur for each wavefunction over mutually exclusive subsets of the spatial domain. This has not been frequently observed in the literature for cubic complex GL systems, as self-interaction and cross-interaction terms are often take to be the same. In the cubic GL system limit ($\mu =0$), we observe the same behaviors for large enough XPM parameters, suggesting that these dynamics occur in appropriate complex GL systems generically, independent of the precise form of nonlinearity so long as the nonlinearity includes self- and cross-interactions, and provided that cross-interactions (XPM parameters in our model) are sufficiently dominant.

We observe \textit{amplitude death} of one wavefunction when there is sufficient asymmetry in the XPM parameters. Indeed, while large enough values of $\alpha_1 = \alpha_2$ promotes a segregation of the individual wavefunctions, we find amplitude death of one wavefunction when $\alpha_1$ and $\alpha_2$ are large and also take different values; see Fig. \ref{fig4}. Amplitude death due to delayed coupling in GL systems has been observed recently by \cite{teki2017amplitude}, where it was shown that amplitude death can occur in a pair of one-dimensional CGL systems coupled by diffusive connections. The analytical results of \cite{teki2017amplitude} reveal that amplitude death never occurs in a pair of identical complex cubic GL systems coupled by a ``no-delay connection", but can occur in the case of the delay connection considered. In our work, using XPM parameters different from unity and obeying certain parameter restrictions, we observe such amplitude death in perhaps a simpler or more natural way than introducing delayed couplings. The results of \cite{teki2017amplitude} stating that coupled GL equations with equivalent kinetics do not have amplitude death hold true for our saturable model as well, as this corresponds to the case of $\alpha_1 = \alpha_2 =1$ in our model, which never satisfies the parameter restrictions \eqref{die1}-\eqref{die2}. Amplitude death or partial amplitude death is more commonly seen in literature on network or lattice equations such as oscillator systems \cite{dodla2004phase,mehta2006death,liu2012effects,nakao2014complex,banerjee2015mean}, and is seemingly novel in reaction-diffusion PDE systems. As was shown recently in \cite{van2018generic}, this amplitude death is not specific to saturable or cubic GL systems, but can be considered a generic behavior for some parameter regimes, provided there is sufficient asymmetry in the self- and cross- coupling terms present in the nonlinearity. See \cite{van2018generic} for more details.

The level of saturation present (as measured by increasing the parameter $\mu$) can regularize or stabilize dynamics in many cases, and can also change the timescale of the dynamics. While we have not commented much on the role of $n$ in the saturable terms, we note that an increase in $n$ will result in a stabilization of the dynamics and a slowing of the timescale. These results suggest that one might modify the saturability of the underlying media in order to develop a novel control mechanism for dynamics from complex GL equations and systems, which has apparently not been considered in the earlier mentioned control literature. Additionally, one could attempt to modify the XPM parameters in order to better control the placement of wavefunctions or force amplitude death if that is desirable. Used in conjunction, saturability and XPM parameters could be used to position and reinforce the intensity of specific patterns. Alternative manners of control could involve the inclusion of spatial or temporal coefficients, which, if properly tuned, could be used to control the wave envelopes through dispersion management \cite{serkin2000soliton}, or the speed of traveling fronts through wavespeed management \cite{baines2018soliton}. Such an approach has previously been applied to the cubic complex GL equation \cite{fewo2005dispersion}, and it may be interesting to consider how such an approach may be used in conjunction with saturability of the media in order to modify the behavior of a nonlinear wave solution to \eqref{sgl1}-\eqref{sgl2}.

While often regularizing, we observe that saturable terms can prolong the existence of transient defects, such as those found in \cite{hernandez1999spatiotemporal}; see Fig. \ref{fig9}. Furthermore, there are regimes where saturability permits blow-up of solutions yet no such blow-up occurs in the cubic nonlinearity case of $\mu =0$. This suggests that the dynamics are highly sensitive to the model parameters, even in the presence of strong saturation in the media. It was interesting to observe what appear to be rogue waves, even when saturation was very large (see Fig. \ref{fig7}). Calculating the rogue wave envelope analytically is made more challenging due to the saturable nonlinearity, and so we only have numerical indications of the emergence of rogue waves. We are not aware of the emergence of rogue waves in saturable models, although there have been experimental suggestions of rogue waves in saturable media. Optical rogue waves were previously reported in complex cubic scalar GL equations \cite{gibson2016optical} with higher-order derivatives, and the emergence of rogue waves in certain parameter regimes of our model for weak saturability is therefore not surprising. We also mention \cite{pierangeli2016turbulent}, which found candidate rogue waves numerically in a nonlinear Schr\"odinger equation with saturable nonlinearity. Like what we have shown for our solutions in Fig. \ref{fig7}, \cite{pierangeli2016turbulent} found isolated large-amplitude waves surrounded by background waves of much smaller mean wave height. A more systematic study on how saturability of the media might influence the stability of highly localized structures, such as rogue waves or breather-like solutions, would be an interesting topic for future work.





\end{document}